\documentclass[aps,pra,floatfix,nofootinbib,reprint,letterpaper,onecolumn,superscriptaddress,longbibliography]{revtex4-2}
\usepackage[left=1.1in,right=1.1in,top=1in,bottom=1in]{geometry}
\usepackage{parskip}

\usepackage{dsfont}
\usepackage{braket}
\usepackage{siunitx}
\usepackage{amsthm}
\usepackage{amsmath}
\usepackage{amsfonts}
\usepackage{physics}
\usepackage{amssymb}
\usepackage{graphicx}
\usepackage{color}
\usepackage{xcolor}
\usepackage{accents}
\usepackage{bbold}
\usepackage[colorlinks=true, citecolor=blue]{hyperref}
\usepackage[capitalise,nameinlink]{cleveref}
\usepackage[normalem]{ulem}
\usepackage[shortlabels]{enumitem}
\usepackage{subcaption} 
\usepackage{listings}
\usepackage{natbib}
\usepackage{mathtools}
\usepackage{float}
\usepackage{changepage}

\DeclarePairedDelimiter\ceil{\lceil}{\rceil}
\DeclarePairedDelimiter\parens{\lparen}{\rparen}

\newcommand{\cB}{\mathcal{B}}
\newcommand{\cW}{\mathcal{W}}

\newcommand{\Rz}{R_z}

\DeclareMathOperator{\ZPow}{ZPow}

\DeclareMathOperator{\QVR}{QVR}
\newcommand{\bgrad}{{b_\text{grad}}}

\makeatletter 
\newsavebox{\@brx}
\newcommand{\llangle}[1][]{\savebox{\@brx}{\(\m@th{#1\langle}\)}%
  \mathopen{\copy\@brx\kern-0.5\wd\@brx\usebox{\@brx}}}
\newcommand{\rrangle}[1][]{\savebox{\@brx}{\(\m@th{#1\rangle}\)}%
  \mathclose{\copy\@brx\kern-0.5\wd\@brx\usebox{\@brx}}}
\makeatother

\theoremstyle{definition}

\theoremstyle{remark}

\definecolor{codegreen}{rgb}{0,0.6,0}
\definecolor{codegray}{rgb}{0.5,0.5,0.5}
\definecolor{codepurple}{rgb}{0.58,0,0.82}
\definecolor{backcolour}{rgb}{0.95,0.95,0.92}
\lstdefinestyle{mystyle}{
  backgroundcolor=\color{backcolour},
  commentstyle=\color{codegreen},
  keywordstyle=\color{magenta},
  numberstyle=\tiny\color{codegray},
  stringstyle=\color{codepurple},
  basicstyle=\ttfamily\footnotesize\color{magenta},
  breakatwhitespace=false,         
  breaklines=true,                 
  captionpos=b,                    
  keepspaces=true,                 
  showspaces=false,                
  showstringspaces=false,
  showtabs=false,
  xleftmargin=0.02\textwidth,
  rulecolor=\color[RGB]{200,200,200},
  frame=bt,
  framextopmargin=2pt,
  framexbottommargin=2pt,
  framexleftmargin=10pt,
  tabsize=2
}
\newcommand{\fig}[1]{\hyperref[fig:#1]{Figure~\ref*{fig:#1}}}
\newcommand{\tbl}[1]{\hyperref[tbl:#1]{Table~\ref*{tbl:#1}}}
\newcommand{\eq}[1]{\hyperref[eq:#1]{Equation~\ref*{eq:#1}}}
\newcommand{\sect}[1]{\hyperref[sec:#1]{Section~\ref*{sec:#1}}}

\begin{document}
\title{Expressing and Analyzing Quantum Algorithms with Qualtran}

\author{Matthew P. Harrigan}
\email{mpharrigan@google.com}
\thanks{Equal contribution}
\affiliation{Google Quantum AI, Venice, CA 90291, United States}

\author{Tanuj Khattar}
\email{tanujkhattar@google.com}
\thanks{Equal contribution}
\affiliation{Google Quantum AI, Venice, CA 90291, United States}

\author{Charles Yuan}
\affiliation{Google Quantum AI, Venice, CA 90291, United States}
\affiliation{MIT CSAIL, Cambridge, MA 02139, United States}

\author{Anurudh Peduri}
\affiliation{Google Quantum AI, Venice, CA 90291, United States}
\affiliation{Ruhr University Bochum, 44801 Bochum, Germany}

\author{Noureldin Yosri}
\affiliation{Google Quantum AI, Venice, CA 90291, United States}

\author{Fionn D. Malone}
\affiliation{Google Quantum AI, Venice, CA 90291, United States}

\author{Ryan Babbush}
\affiliation{Google Quantum AI, Venice, CA 90291, United States}

\author{Nicholas C. Rubin}
\email[]{nickrubin@google.com}
\affiliation{Google Quantum AI, Venice, CA 90291, United States}

\begin{abstract}

Quantum computing's transition from theory to reality has spurred the need for novel software tools
to manage the increasing complexity, sophistication, toil, and fallibility of quantum algorithm
development. We present Qualtran, an open-source library for representing and analyzing quantum
algorithms. Using appropriate abstractions and data structures, we can simulate and test algorithms,
automatically generate information-rich diagrams, and tabulate resource requirements. Qualtran
offers a \emph{standard library} of algorithmic building blocks that are essential for modern
cost-minimizing compilations. Its capabilities are showcased through the re-analysis of key
algorithms in Hamiltonian simulation, chemistry, and cryptography. Architecture-independent resource
counts output by Qualtran can be forwarded to our implementation of cost models to estimate physical
costs like wall-clock time and number of physical qubits assuming a surface-code architecture.
Qualtran provides a foundation for explicit constructions and reproducible analysis, fostering
greater collaboration within the growing quantum algorithm development community.

\end{abstract}

\maketitle


\section{Introduction}

As quantum computing technology develops so too must our ability to leverage it for computational advantages. Luckily, we know of many applications for which quantum algorithms can give a meaningful asymptotic speedup relative to a classical computer~\cite{grover1996fast, shor-1994, PhysRevLett.83.5162, ambainis2007quantum, jordanalgzoo}. The real-world computational advantage of a quantum algorithm, however, depends not only on its idealized asymptotic time complexity, but also on constants governed by algorithmic implementation and physical realities of quantum architectures such as error correction. Notably, the most competitive forms of quantum error correction known today incur large space-time overheads that can jeopardize advantage~\cite{Babbush_2021,hoefler2023}.
Thus, accurately determining the advantage of a quantum algorithm running on hardware requires fully compiling the algorithm and determining its cost in terms of concrete constants. While resource estimation is not a new activity in quantum information~\cite{PhysRevA.79.062314} the scale and complexity of modern quantum algorithms demands improved tools for algorithm analysis and construction.

To analyze the constant factors of compiling a quantum algorithm to an error-corrected quantum computer, researchers typically record each operation using a mixture of enumerated lists and tables, mathematical Dirac notation, and English prose in a scientific journal article~\cite{Babbush2018Encoding, rubin2024quantum, Sanders_2020,Su2021FirstQuant,vonburg2021catalysis, zini2023quantum, OBrien2022, berry2023quantum}. 
The challenge is that state-of-the-art quantum algorithms exist at the scale of \emph{teraquops} --- $10^{12}$ qubits $\times$ operations~\cite{Lee2021Even}. 

At that scale, a manual analysis is tedious, consuming months of researcher time and dozens of pages in a publication.
For example, the manual resource analysis in Ref.~\cite{Sanders_2020} consumes Appendices A--D (16 pages), in Ref.~\cite{Lee2021Even} it consumes Appendices A--F (37 pages), and in Ref.~\cite{vonburg2021catalysis} it consumes Appendix VII (41 pages).
Such analysis is imprecise, requiring a researcher to approximate in an \textit{ad hoc} way the cost
of many operations lacking clear references from the literature. It is also inaccessible, yielding
numerical figures that are hard for a reader to reproduce or extend.

Researchers have previously developed quantum programming languages, including
Quipper~\cite{green2013quipper}, Scaffold~\cite{javadiabhari2015scaffcc}, QWire~\cite{paykin2017},
Q\#~\cite{svore2018q}, Silq~\cite{bichsel2020}, Tower~\cite{Yuan_2022}, and Qrisp~\cite{seidel2024},
that enable the programmer to express algorithms in a structured way that respects the abilities and
constraints of a theoretical quantum computer. There are also existing frameworks for programming
noisy intermediate-scale quantum (NISQ) computers at the circuit level within a Python host language
such as Cirq~\cite{cirq_developers_2023_10247207}, pyQuil~\cite{smith2016practical}, and
Qiskit~\cite{Qiskit}. Taking inspiration from the useful features of
existing languages, we contend that there is ample opportunity to experiment with new abstractions,
tools, and ideas for quantum programming. 

\begin{figure}[t]
    \centering
    \includegraphics[width=1.0\linewidth]{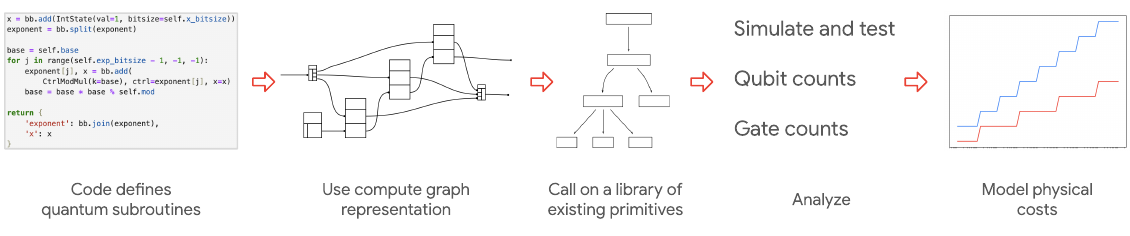}
    \caption{A cartoon overview of a Qualtran workflow. Python code defines quantum subroutines and
    properties. The data structures encode quantum invariants. Routines are composed hierarchically,
    and Qualtran provides a library of useful quantum subroutines. Analysis protocols let the
    developer make meaningful statements about the correctness or cost of a quantum algorithm.
    Architecture-agnostic costs are translated to physical resource requirements like wall-clock
    time through standard models.}
    \label{fig:toc}
\end{figure}

In this manuscript, we introduce Qualtran, an open-source, Python-based software framework that
enables a researcher to express and analyze quantum algorithms for error-corrected quantum computers
in an automated, precise, and accessible way. Qualtran offers 
a) data structures to represent quantum programs by hierarchically composing subroutines;
b) methods to analyze the performance of quantum programs, including hardware resource costs; 
c) a library of re-usable algorithms and subroutines based on recent research on compilation and resource estimation; and 
d) a set of hardware architecture-aware models for estimating physical costs from architecture-agnostic costs.

We begin with a description of the design of the software tool---its data structures and analysis
routines---in \cref{sec:design}. We then provide a survey of algorithmic primitives we consider
essential for modern compilation studies in \cref{sec:primitives}. Next,
\cref{sec:hamiltonian,sec:ground-states,sec:cryptography} demonstrate how Qualtran can be used on
representative problems in quantum algorithms compilation including Hamiltonian simulation, ground
state chemistry problems, and breaking cryptosystems. We describe the various physical cost models
for translating the hardware-agnostic logical costs analyzed in the prior sections to physical
quantities in \cref{sec:hardware-costs}, and conclude with closing remarks in
\cref{sec:conclusion}.

\section{Design of the Qualtran Framework\label{sec:design}}

In designing the Qualtran framework, we adopted design principles that we believe distinguishes it
from prior work. In particular, we have aimed to explicitly address the needs of practicing quantum
algorithms researchers, who are already publishing studies on compiling large quantum
algorithms---often without the use of any software tool. We specifically aim to accelerate this work rather than obviate it.

One principle is that the framework itself should be accessible and predictable. We motivate this by
a contrast with classical programming languages: the C++ programming language has hundreds of
developers contributing to the language infrastructure and millions of programmers agnostic to the
particulars of the compilation infrastructure. Quantum algorithms research is not at this stage of
maturity. Instead, researchers have to understand the problem domain, quantum algorithms, techniques
for compiling quantum algorithms, and analysis techniques for reasoning about the cost and
correctness of a program one cannot yet run. We designed Qualtran so that the \emph{runtime}---in
addition to the library of quantum algorithms---is readable, documented, extensible Python code. The
software tries to be as predictable as possible: there is no special syntax, mutable state is
avoided or carefully controlled, interfaces and types are explicit, and interoperability with the
existing ecosystem of scientific computing Python packages is straightforward. 

A second principle is that researchers need only annotate as much---or as little---about their
algorithms and subroutines as is necessary to complete their research objectives. We seek meaningful
statements about the correctness or cost of a quantum algorithm, not an executable `binary' for a
quantum program. When authoring a quantum subroutine, the user can either employ a placeholder with
literature-based cost expressions or provide a full implementation of the subroutine. This
flexibility is powerful: the system can emit a compiled circuit composed of logical 1- and 2-qubit
gates if full implementations of each subroutine is provided. Alternatively, a user can quickly
tabulate resource costs by outlining the expensive part of an algorithm.

The final principle is that the \emph{standard library} of quantum algorithm primitives included
in the framework should prioritize subroutines developed for---and used by---recent resource
estimation studies. \Cref{sec:primitives} provides additional detail for this principle, and
highlights some of the most important primitives available in the standard library.

In the remainder of this section, we give an overview of the design of the Qualtran framework and
how it enables a developer to express and analyze quantum algorithms. We first introduce the central
data structures in Qualtran that enable a developer to build quantum programs in a modular way. We
then present protocols for analyzing quantum programs in Qualtran, including gate counting, qubit counting,
and simulation.

\subsection{Bloqs: Building Blocks of Quantum Algorithms}

To enable the developer to express quantum algorithms, Qualtran provides \emph{bloqs}, which are
data structures that represent quantum operations, subroutines, and programs. Some examples of
bloqs include simple unitary quantum logic gates over individual qubits such as Hadamard, phase, and
multi-controlled $\mathrm{NOT}$ gates; complex operations acting on registers of qubits such as
multiplication, reflection, and state preparation; and select non-unitary operations including qubit
allocation and deallocation.

We introduce a new term for this category of objects to avoid any implied constraints imposed by
existing labels, and chose the name \emph{bloq} to highlight these objects' role as the building
blocks of a quantum algorithm. A minimal bloq is characterized only by its name and
\emph{signature}---the names and dimensions of its input and output quantum registers. All other
properties, such as its underlying sequence of logic gates or its numerical denotation as a tensor,
are optional annotations on a bloq. This optionality enables a developer to implement only the
properties of a bloq for the analysis they desire to perform. For example, annotating the
approximate gate count for the most costly subroutine in an algorithm can be used for resource
estimation. With additional detail, the framework enables compilation of a program down to an
elementary gate set.

In contrast to classical programming languages and many quantum languages, Qualtran does not
represent quantum programs as a sequence of textual keywords and identifiers. Instead, we adopt a
model in which bloqs are represented explicitly as data structures within the Python host language.
This model is common in NISQ programming
frameworks~\cite{cirq_developers_2023_10247207,smith2016practical,Qiskit}, as well as machine
learning frameworks such as TensorFlow~\cite{tensorflow2015}. This approach provides us access to a
general-purpose programming language for analyzing, visualizing, and testing algorithms, and we
believe this combination of a familiar host language with abstractions for quantum algorithmic
building blocks is particularly accessible and useful for researchers in quantum algorithms.

\subsubsection{Composition of Bloqs}

\begin{figure}
  \includegraphics[width=0.75\textwidth]{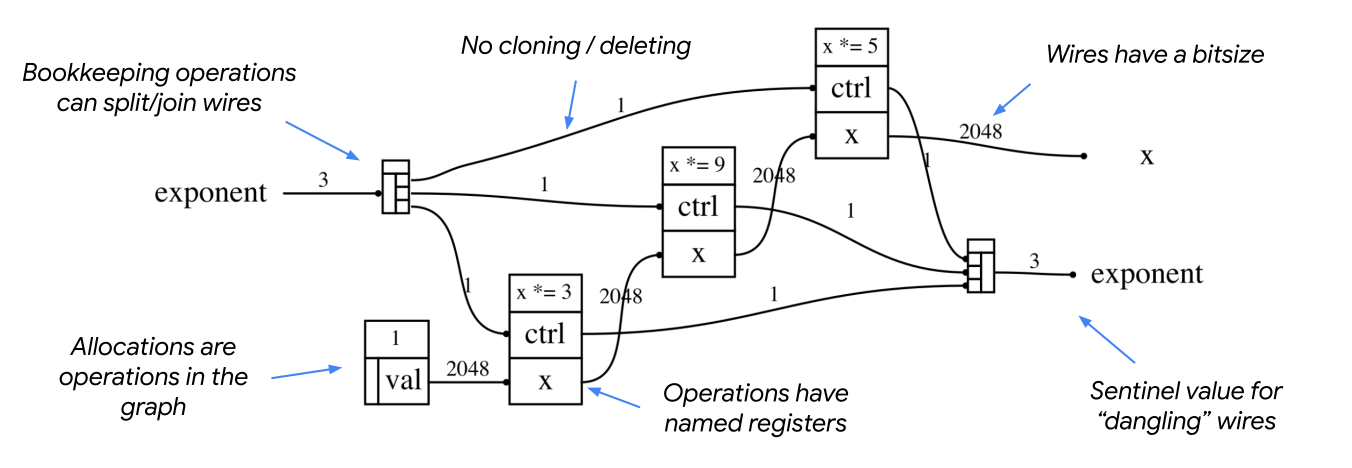}
  \caption{The compute graph for a modular exponentiation subroutine. This graph
  is formed by composing modular multiplication bloqs and serves as the definition of
  a modular exponentiation bloq. Some notable features of the data structure are annotated.
  \label{fig:cbloq-features}
  }
\end{figure}

In Qualtran, the user defines a bloq by its decomposition 
into simpler bloqs. Qualtran provides atomic bloqs, such as individual logic
gates, that serve as the base case of this recursive strategy. 
For any bloq, its constituents composed to 
form a \emph{compute graph}---a directed acyclic graph (DAG) whose edges, or \emph{wires},
represent the flow of quantum data.

In \cref{fig:cbloq-features}, we illustrate the compute graph formed by composing modular
multiplication bloqs to define a modular exponentiation bloq. We can observe the graph
structure and trace the flow of quantum information from left to right. Each edge represents the
flow of quantum data from the output of one bloq to the input of a subsequent bloq. A wire may
represent one qubit or a larger (quantum) value with a particular \emph{quantum data type} such as
an unsigned integer encoded in multiple qubits. In the example, the \lstinline{Allocate} bloq outputs a 2048-bit quantum integer to accumulate the results of the subsequent multiplications. A user that needs fine-grained
manipulation of the encoded data can use the \lstinline{Split} bloq to address
individual qubits, as seen in the treatment of the exponent register. Each qubit in the exponent is used 
as a control for a corresponding multiplication operation.

These two features, hierarchical decomposition of bloqs and quantum data types, improve the classical runtime
performance of the library. By incrementally lowering one level of the bloq hierarchy at a time,
each decomposition step need only concern itself with tens or thousands of constituent bloqs. It does not need to fully unroll
a bloq into e.g. all $10^9$ gates that would be needed to represent the bloq as a single circuit.
Additionally, the classical memory footprint of the Python object for the bloq in
\cref{fig:cbloq-features} would not increase if the bit-width of the arithmetic was increased from
2048 to 4096 bits, since the individual bits of a large quantum data type need not be explicitly represented.

This design also offers correctness guarantees. For example, the no-cloning~\cite{Wootters1982ASQ}
and no-deletion~\cite{Pati2000} theorems mandate that every piece of quantum data must be used
exactly once in any physically realizable quantum computation. Qualtran enforces this correctness
property automatically and transparently to the user via principles of linear logic developed in
programming language research~\cite{paykin2017}. The mechanism is that, under the hood, the user does
not manage the graph structure directly but rather invokes user-facing functions to manipulate the
wire connection points. These functions raise a runtime error whenever a wire connection point is
used twice, or is left unused without being marked as a bloq's input or output.

Finally, Qualtran avoids a class of errors due to misidentifying qubits~\cite{paltenghi2022} by
eschewing the concept of an implicit pool of qubits or qubit addresses as found in existing
frameworks~\cite{cirq_developers_2023_10247207,smith2016practical,Qiskit}. Allocations and
de-allocations are explicit bloqs in the compute graph. Also, each register has a string name used during composition
rather than relying on order or index alone.

\subsubsection{Quantum Data Types}

The elementary unit of quantum information is the qubit. We can use a collection of qubits to encode data with
\emph{quantum data types}---such as integers and real numbers up to a given precision. As introduced
in the previous section, Qualtran offers first-class support for quantum data types. The signature
of a bloq reports not only the arity and names of its inputs and outputs, but also its quantum data
type. Bloqs can only be composed if their wires have matching data types. Each data type can be
queried to determine how the data is encoded to and from (qu)bits. The number of qubits for a data
type is crucial for use in resource estimation.

Whereas classical programming languages have a fixed collection of primitive types based on the
machine architecture such as \lstinline{int32} or \lstinline{float64}, the quantum data types in
Qualtran enable the user to specify an arbitrary bit-width for a register. For example, the
\lstinline{QInt(8)} type specifies an 8-bit signed integer and the \lstinline{QFxp(6, 4)} type
specifies a 6-bit fixed-point number with 4 fractional bits. This flexibility is important for estimating
resource requirements of quantum algorithms that we anticipate to be limited by qubit count, and
where using a larger-than-necessary type for a known range of values is too wasteful for emerging
hardware architectures.

Qualtran provides data types for unsigned integers, one's and two's complement signed integers,
integers in Montgomery form~\cite{Montgomery1985}, and fixed-point real numbers. The framework
enables the user to define custom data types by implementing the appropriate interface. Additionally,
it offers special support for fixed-size multidimensional arrays of quantum data. For example, a bloq
that selects a register from $m$ input registers and swaps it into the first position can define its
signature as one control register and an array of $m$ \lstinline{QInt} registers.

Qualtran does not support quantum sum (union) or product (struct) types, and by extension does not
support quantum strings or quantum lists~\cite{Yuan_2022}. This design accounts for the relative
difficulty of estimating the resource costs of such data types under constraints of foreseeable
quantum architectures, but may be revisited in future versions of the library.

\subsection{Protocols for Analyzing Quantum Programs}

We cannot yet execute the large-scale quantum algorithms expressed in Qualtran that require an error-corrected quantum computer.
Nevertheless, researchers today labor to prototype, compile, and analyze quantum algorithms so that we may accurately predict the practical speedup of a given algorithm and obtain the most use out of an emerging hardware device. 
Qualtran aids researchers in this goal through protocols that act on quantum programs built from bloqs to query or derive useful properties, produce visualizations and diagrams, and check correctness of algorithms and subroutines.

\subsubsection{Call Graphs}

A basic task given a quantum program in bloq form is to compute and visualize the set of bloqs to which it decomposes, which in turn empowers downstream analyses such as gate and qubit counting. 
The data structure that Qualtran uses for this analysis is a \emph{call graph}, a directed acyclic graph that has an edge pointing from a \emph{caller} bloq to a \emph{callee} bloq whenever the callee is in the collection of bloqs to which the caller decomposes. Each edge is annotated with the number of times the callee is called.
Qualtran represents a call graph as a NetworkX~\cite{hagberg2008} graph obtained by calling a designated method on a bloq. 

A bloq can define its behavior and callees in two ways. The first way is to define a \emph{full decomposition} that specifies the compute graph, including the explicit flow of information between the inputs and outputs of callees. In turn, Qualtran recovers a call graph from a compute graph by discarding detail and retaining only relationships between callers and callees. A full decomposition is needed for tasks such as simulation, but is not always required to accomplish a researcher's goals when considering the effort. 

As an alternative, the bloq may define a \emph{call graph decomposition} that simply lists the callees (and their number of calls) without accounting for the flow of quantum information. This approach is sufficient for certain analyses, such as gate counts, and has three advantages over full decomposition:

\begin{itemize}
\item The user can quickly estimate costs by only providing the unstructured list of callees. 
\item Parameters can propagate as symbolic rather than concrete values. 
For example, consider a bloq that applies the $X$ gate on a parametric number of qubits $n$. Its full decomposition would instantiate $n$ \lstinline{XGate} objects, and the Python runtime cannot instantiate a symbolic number of objects. By contrast, the call graph decomposition simply states that the bloq uses $n\ X$ gates, where $n$ is a SymPy~\cite{sympy} variable that supports arithmetic.
\item The call graph decomposition can serve as an automated cross-check of correctness against a full decomposition of the bloq if the user provides both.
\end{itemize}

Given a bloq, we can visualize its compute and call graphs in order to validate constructions and communicate algorithms.
Qualtran supports drawing these graphs, complete with annotations on each bloq such as qubit and gate counts. The framework can draw the compute graph of a bloq either as a directed acyclic graph as in \Cref{fig:cbloq-features} via Graphviz~\cite{graphviz}, or as a traditional quantum circuit via $\bra{\mathsf{q}}\ket{\mathsf{pic}}$~\cite{qpic}.\footnote{In Qualtran, we term this style a \emph{musical score} diagram for disambiguation from other references to circuits.}

\subsubsection{Gate Counting}

The first essential question about a quantum algorithm is how many operations must be performed to
execute the algorithm. Although the associated physical wall-clock time will depend on details of
the target hardware architecture and error-correcting code (see \cref{sec:hardware-costs}), the
quantity and type of quantum logic gates serves as a useful architecture-independent proxy for the
time cost of an algorithm.

Qualtran enables a researcher to automatically count the number of gates comprising a quantum
algorithm. For an algorithm expressed as a bloq, the system recursively sums the contributions from
the callees to which the bloq decomposes. This analysis is simple and is not fundamentally tied to
any particular target architecture. The analysis is also efficient: by automatically caching the
computed costs of reoccurring bloqs, Qualtran can compute gate counts for algorithms containing
hundreds of millions of gates in a matter of seconds.

There are two specific strategies included for gate counting. The \lstinline{BloqCounts} strategy
requires researchers to explicitly specify the bloqs they wish to count, but affords the user
complete control over the target gateset or architecture. In contemporary literature on quantum
resource estimation, however, researchers often map an algorithm onto a quantum architecture that
uses the surface code for error correction. To account for gate costs, researchers typically count
$T$ and Toffoli gates, the quantity (but not specific identity) of Clifford gates, arbitrary-angle
rotation gates, and measurement operations
\cite{Horsman2011SurfaceCQ,Gidney2019efficientmagicstate,beverland2022assessing}. The
\lstinline{QECGatesCost} strategy that is pre-configured to count the gates most relevant to a
surface code error corrected architecture.

In designing this second strategy, it became apparent that there is no strict agreement in the
literature on \emph{specifically} which fault tolerant gates to count and when to `stop' decomposing. Some studies
will continue compiling an algorithm until only Clifford and $T$ gates remain; while others will
count other basic gates like the Toffoli or Fredkin gate independently of the $T$ count. Indeed,
there are some surface code architectures that have different costs for directly implementing
Toffoli gates compared to compiling them to four or seven $T$ gates.
In our design of the \lstinline{QECGatesCost} gate counting routine, we keep the counts
\emph{factored} by their gate type as much as possible and provide methods to aggregate 
counts for a particular choice of gate set.

The gate counts can be insightful in their own right by comparing the runtime costs of algorithms in an
architecture-independent way, or they can be fed into the physical cost models detailed in \cref{sec:hardware-costs}.

\subsubsection{Qubit Counting}

The second essential property of a quantum algorithm is how many qubits it requires. While the
number of required physical qubits will also depend on the target hardware architecture (see
\cref{sec:hardware-costs}), the number of logical, error-corrected qubits is
architecture-independent. Using Qualtran, a researcher can automatically estimate the number of
logical qubits required by a quantum algorithm.

Qualtran counts qubits in a hierarchical and recursive way. It assumes that each callee in the
decomposition of a bloq executes sequentially, and computes the number of qubits at each sequence
point as the number of qubits required by the callee plus the number of idle qubits acting as
bystanders. In assuming that callees execute sequentially, Qualtran minimizes qubit count
potentially at the expense of greater circuit depth or execution time. Conversely, by assuming that
each callee uses all of its qubits for its entire execution, Qualtran potentially overestimates the
qubit count. We believe that this Min-Max-style estimate can provide a good balance between accuracy
and scalability of the accounting. 

We discuss two important considerations for designing a qubit counting algorithm: 
First, bloqs can allocate and free qubits within their decomposition; so we cannot
simply look at the input and output sizes of a bloq to determine its qubit cost. Second, the qubit count
cannot be expressed as the maximum over the qubit counts of all the callees: when executing a particular
callee, the number of qubits is at least the number of qubits required by the callee \emph{plus} the 
number of idle, bystander qubits that are output or inputs to other callees.

There are optimizations that are only possible by fully decomposing a bloq to a circuit by
flattening hierarchical structure and carefully tessellating operations from different components.
Such strategies are employed for circuits of thousands of gates that run on current generation, NISQ
hardware. Indeed, one can use the interoperability features of Qualtran to convert a program to a
Cirq circuit with such optimizations, but this is likely not feasible for large-scale algorithms
with billions of gates. It is a future direction to investigate more accurate qubit allocations
without sacrificing tractability. 

Overall, Qualtran makes simplifying assumptions by default that efficiently estimate the number of
qubits used by many quantum algorithms, and we have designed the the framework to be sufficiently
general that a developer can write their own routine for counting qubits if additional precision is
necessary. 

\subsubsection{Tensor Simulation}

In the absence of a hardware quantum computer, researchers often check whether a quantum program is correct by simulating it on a classical computer. In general, such simulation requires an exponential amount of classical resources. A state vector simulation stores $2^n$ complex numbers to simulate an $n$-qubit system, while a Feynman path integral simulation uses constant space but exponential time to sum over an exponential number of paths~\cite{bv1997}. Thus, simulation is tractable only for bloqs that use few qubits or exhibit special structure. Nevertheless, by testing the numerical simulation of small instances of a bloq, we gain confidence that the implementation is correct at problem-relevant scales as well.

Qualtran can simulate a quantum program by deriving its state vector or unitary (its \emph{tensor})
from the tensors of its constituent bloqs. 
We use the tensor network techniques provided by Quimb~\cite{gray2018quimb} to find good tensor contraction orderings---which interpolate between the state vector and Feynman path approaches---that decrease the cost of simulating complex bloqs. As an example, we have successfully used Qualtran to test that a 32-bit quantum adder circuit consisting of 64 qubits works correctly on given input/output states via a simulation that uses only 6 qubits' worth of classical memory. 

\subsubsection{Classical Simulation\label{sec:classical-sim}}

Whereas full tensor simulation has exponentially large cost in general, many common quantum algorithmic primitives merely perform classical reversible logic on data that just happens to exist in superposition. We can efficiently simulate such operations as part of testing a complex algorithm on classical inputs and outputs. For example, the \emph{quantum lookup table} primitive of \Cref{sec:quantum_lookup_table} has semantics identical to a table lookup on a classical computer, except that the index register contains a superposition of indices so that the operation loads a superposition of the classical data at each index.

Qualtran has explicit support for simulating and testing bloqs that encode classical logic as above. First, a bloq author annotates how the bloq acts over classical data by providing corresponding Python code. Qualtran uses this logic to propagate classical values through bloqs and produce a classical output. 
Then, the user can perform an analogue to fuzz testing in classical software engineering by sampling a set of classical input/output pairs and comparing to the outputs of the classical simulation. 
After testing enough pairs of classical values, the user can gain confidence that the quantum construction is correct.

\section{Essential Quantum Subroutines in the Qualtran Standard Library\label{sec:primitives}}

In this section, we present a core set of subroutines and techniques to manipulate quantum information that underpin the diverse world of quantum algorithms. These techniques are not traditionally taught as part of quantum information classes, nor do they appear in standard quantum computing textbooks. Instead, they emerge from quantum resource estimation studies from the last decade, painstakingly derived again and again in the prose of paper appendices. 
By collecting these subroutines in the Qualtran standard library, we aim to bring these essential tools to the forefront and empower researchers to quickly express quantum algorithms at a high level of abstraction without sacrificing the latest advancements in performance engineering.
However, we view this library as merely the beginning, not the end. Our aspiration is to vastly expand it over time as we continue our work on resource estimation. We also hope that the open source community will actively contribute to building more primitives for the standard library, fostering a collaborative effort to enrich this vital resource for quantum algorithm development.

In each of the following sections, we highlight the quantum algorithmic primitive as a bloq or set of bloqs and demonstrate its construction through code snippets along with error propagation and visualization:
\begin{itemize}
\item In \Cref{sec:rotations}, we describe Qualtran's features for performing rotations of quantum states that are used by modern fault-tolerant algorithms. Supported rotations include direct synthesis, phase kickback on a Fourier state, and probabilistic implementations using generalized state injection.
\item In \Cref{sec:unary-iteration}, we describe Qualtran's features for selecting quantum operations via a quantum index using iterator circuits that can be reused in a wide variety of tasks. Specifically, in \Cref{sec:quantum_lookup_table}, we show how this tool can be used to load classical data into quantum superposition.
\item In \Cref{sec:quantum_state_prep}, we describe Qualtran's features for preparing arbitrary quantum states and arbitrary purified density matrices. We describe the context in which these two state preparation tasks are useful and highlight the strategies supported by Qualtran for dense and sparse versions of these tasks. 
\item In \Cref{sec:block-encoding}, we describe Qualtran's block encoding library, which enables quantum algorithms to load datasets or transformations of states as matrices and to perform arithmetic operations such as tensor products, products, and linear combinations on the encoded matrices.
\item In \Cref{sec:qsp}, we describe Qualtran's support for quantum signal processing, which enable algorithms to transform block encodings via arbitrary polynomials.
\item Finally, in \Cref{sec:phase-estimation}, we present bloq constructions of modern phase estimation techniques using window functions.
\end{itemize}

\subsection{Rotations}\label{sec:rotations}

Rotations are a fundamental class of operations in quantum algorithms with substantial space/time cost differences depending on their implementation. The selection of an optimal strategy to perform arbitrary rotations can depend on algorithmic context and hardware context---i.e. limiting regimes such as early fault-tolerant quantum computers with few logical qubits. Thus, Qualtran provides a number of ways to achieve arbitrary rotations to varying precision with different quantum resource overheads. While we do not support program analysis to automatically choose the optimal rotation primitive, many bloqs in Qualtran default to one of the following strategies which make the most algorithmic sense.

The lowest level rotation primitives supported by Qualtran are single-qubit $Z$-axis rotations. 
We consider two standard conventions defining the rotation, $\Rz$ and $\ZPow$,
\begin{equation*}
  \Rz(\theta) 
  = e^{-i Z \theta / 2} 
  = \begin{pmatrix}
    e^{-i \theta/2} & 0 \\ 0 & e^{i \theta/2}
  \end{pmatrix}
  \;\;,\;\;
  \ZPow(t)
  = \begin{pmatrix}
    1 & 0 \\ 0 & e^{i \pi t}
  \end{pmatrix}
  = e^{i\pi t/2} \Rz(\pi t)
\end{equation*}
which are equivalent up to global phase. The parameterization of each gate has different domains; $\Rz$ angle $\theta \in [0, 2\pi)$ satisfying $\Rz(\pi) = -iZ$, and for $\ZPow$ $t \in [0, 2)$, satisfying $\ZPow(1) = Z$.
At the time of publication, the Qualtran standard library supports three techniques to achieve these rotations which are described in \Cref{tbl:rotations} along with their resource costs. 
\begin{table}[H]
\centering
\renewcommand{\arraystretch}{1.5}
\begin{tabular}{|c|c|p{5cm}|}
    \hline
    Strategy & Bloq & Non-Clifford Gate Cost 
    \\
    \hline
    Direct synthesis~\cite{Bocharov_2015}
    & \lstinline|ZPow|, \lstinline|Rz|
    & $\ceil*{1.149 \log(\frac1\epsilon) + 9.2}$ T-gates
    \\
    \hline
    Programmable ancilla~\cite{Cody_Jones_2012}
    & \lstinline|ZPowUsingProgrammedAncilla|
    & \par $n=\lceil\log_{2}(1/p)\rceil$ resource states 
    \par $n \cdot \ceil*{1.149 \log(\frac{n}{\epsilon}) + 9.2}$ T-gates
    \\
    \hline
    Phase-gradient~\cite{Sanders_2020}
    & \lstinline|ZPowConstViaPhaseGradient|
    & $(\bgrad - 2)$ Toffoli gates
    \\
    \hline
\end{tabular}
\caption{Bloqs synthesizing $\epsilon$-approximate single qubit $Z$-rotations and their resource costs.
The direct synthesis approach synthesizes $Z$ rotations via a sequence of Clifford+T gates. The programmable ancilla strategy consumes single-qubit resource states prepared ahead of time, described in Eq.~\eqref{eq:prog_ancilla_rot_def}, to apply $Z$-rotations via a stochastic protocol using gate teleportation. The $p$ parameter in the programmable ancilla strategy is the desired success probability of stochastic protocol and affects the number of resource states $n$ to be consumed. Prior work~\cite{Cody_Jones_2012} used Fowler sequences~\cite{fowler2010constructingarbitrarysteanecode} to prepare the resource states in a chemistry context. In Qualtran, we use the direct synthesis approach as the default way to prepare $n$ resource states with accuracy $\epsilon / n$.
The phase-gradient technique synthesizes $Z$ rotations via addition into a phase gradient register of size $\bgrad = \log\frac{2\pi}{\epsilon}$ qubits, as described in \cref{eq:phase-gradient-state}. Note that the phase gradient register has a one-time preparation cost and can be reused to synthesized multiple rotations.
}
\label{tbl:rotations}
\end{table}

For arbitrary axis single-qubit rotations,
Qualtran provides the \lstinline{SU2RotationGate} which accepts any $2 \times 2$ unitary matrix and a precision $\epsilon$, and decomposes into three $Z$ rotations each with precision $\epsilon/3$.
Controlled variants of all single-qubit rotations ($Z$-axis and arbitrary axis) are also supported using circuit identities which reduce controlled Z rotations to uncontrolled Z rotations~\cite{Cody_Jones_2012, nielsenchuang}.

The Qualtran standard library also provides support for a broad class of generalized basis state phasing through the 
\emph{Quantum Variable Rotation (QVR)}~\cite{Cody_Jones_2012,Sanders_2020} set of bloqs. Two examples are, optimized versions of Hamming weight phasing~\cite{Gidney_2018} and phasing by an arbitrary cost functions used in recent resource estimation studies of combinatorial optimization~\cite{Sanders_2020}.

\subsubsection{Single Qubit \textit{Z} Rotation}
\label{subsec:rotations:known-angle}

Here we expand on the single qubit $Z$-axis rotations described in Table~\ref{tbl:rotations} and provide an example invocation code block. 

\medskip
\paragraph{Direct Clifford+T Synthesis.}
\Citet{Bocharov_2015} synthesized $Z$ rotations by the use of \emph{repeat-until-success} circuits
which have an expected T-gate cost of
\begin{equation}
\label{eq:rotation:direct-synth-t-cost}
  \ceil*{1.149 \log \frac1\epsilon + 9.2}.
\end{equation}
The bloqs \lstinline|ZPow| and \lstinline|Rz| %
accept a precision $\epsilon$, and rotation exponent/angle ($t$, $\theta$ respectively). The rotation angle and precision inputs can be invoked as symbolic values or numerical values.   

\medskip
\paragraph{Programmable Ancilla Rotations.}
Originally presented in~\citet[Sec. 2.3]{Cody_Jones_2012}, this technique performs $Z$ rotations  using state-injection into a repeat-until-success circuit. 
Such techniques are often useful in situations where expensive resource states can be prepared ahead of time (possibly in parallel), reducing overhead during the run of the algorithm.
To synthesize an $\ZPow(t)$, this method uses a set of pre-prepared resource states $\ket{\omega^{(k)}}$, called \emph{programmed ancillas}, defined as
\begin{align}\label{eq:prog_ancilla_rot_def}
  \ket{\omega^{(k)}} = \ZPow(2^k t) \ket+ = \frac1{\sqrt2} \parens*{\ket0 + e^{2^k i \pi t}\ket1}.
\end{align}

The bloq's instantiation requires three parameters:
exponent $t$,
precision $\epsilon$,
and the expected number of rounds $n$ which defaults to 2.
The total rotation cost is computed by adding costs to synthesize $n$ resource ancilla in state $\ket{\omega^{(k)}}$, each with precision $\epsilon/n$.

\medskip
\paragraph{Rotation by addition into a phase-gradient state.}
Rotations can be performed by phase kickback when adding a value into 
a \emph{phase-gradient state}~\cite{Kitaev2002ClassicalAQ,Sanders_2020,Gidney_2018}. The phase-gradient state is a Fourier state defined over $b_{\mathrm{grad}}$ qubits
\begin{equation}
\label{eq:phase-gradient-state}
  \ket{\phi_\text{grad}}
  = \frac1{\sqrt{2^\bgrad}} \sum_{k = 0}^{2^\bgrad-1} e^{-2\pi i k/2^\bgrad} \ket{\frac{k}{2^\bgrad}}
  = \bigotimes_{j=1}^{\bgrad} \frac{\ket0 + e^{-2\pi i k/2^j} \ket1}{\sqrt2}
\end{equation}
where $b_{\mathrm{grad}}$ sets the precision for the rotation. Qualtran supports construction of this state using the bloq \lstinline|PhaseGradientState| which can be used as an input wire for the rotation. The phase gradient state $\ket{\phi_\text{grad}}$ is catalytic in the sense that it can be reused for multiple rotations.

\begin{figure}[H]
\centering
  \begin{subfigure}{0.45\textwidth}
  \begin{lstinputlisting}[language=Python, basicstyle=\footnotesize\tt]{snippet_ZPowConstPhaseGradient.py}
  \end{lstinputlisting}
  \end{subfigure}
  \begin{subfigure}{0.45\textwidth}
  \includegraphics[width=\textwidth]{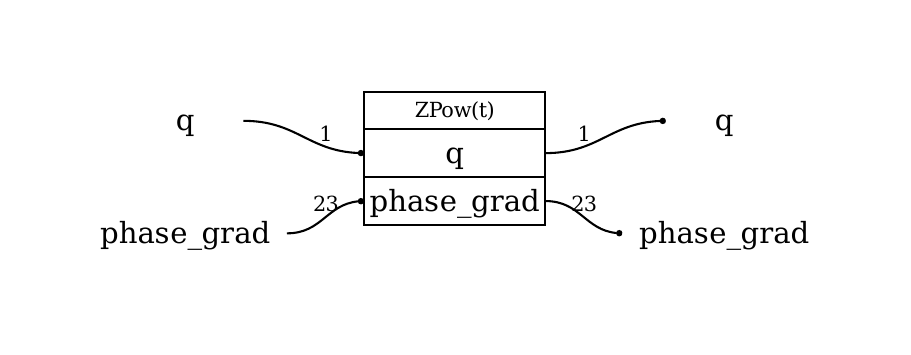}
  \end{subfigure}
  \caption{Demonstration of a code snippet instantiating a $\ZPow(t)$ bloq achieved using a phase-gradient resource state specified to a $1.0\times 10^{-6}$ precision.  The precision is used to dictate the size of the phase-gradient state shown on the right. This bloq has a Tofolli cost of $\bgrad - 2$~\cite{Gidney_2018,Cody_Jones_2012}.}
  \label{fig:rotation:zpow-const-phasegrad}
\end{figure}

The rotation bloq \lstinline|ZPowConstViaPhaseGradient|
accepts an exponent $t$ and the phase-gradient size $\bgrad$,
and implements a $\ZPow(t)$. 
We provide a utility function \lstinline{ZPowConstViaPhaseGradient.from_precision} to allow the user to construct the rotation from precision requirements,
shown in \cref{fig:rotation:zpow-const-phasegrad}.

\subsubsection{Quantum Variable Rotation}
\label{subsubsec:QVR}

A \emph{Quantum Variable Rotation (QVR)}~\cite{Cody_Jones_2012, Sanders_2020}
is a unitary which phases each computational basis state $|x\rangle$ by $e^{2\pi i x}$.
QVR has been used as a primitive for quantum chemistry simulations~\cite{Cody_Jones_2012} and combinatorial optimization~\cite{Sanders_2020}.
An $n$-qubit QVR unitary can be defined as
\begin{equation*}
  \QVR_{n}
  \sum_{j=0}^{2^n-1} c_j\ket{x_j}
  \mapsto
  \sum_{j=0}^{2^n-1} e^{2\pi i x_j} c_j\ket{x_j}
\end{equation*}
Here, the basis $x_j \in [0, 1)$ are interpreted as $n$-bit fractional values.
In general, we implement an approximate $\QVR$, denoted by $\QVR_{n, \epsilon}$,
which is $\epsilon$-close to the above unitary by the distance measure from Ref.~\cite[Eq. 1]{fowler2010constructingarbitrarysteanecode}.

Qualtran provides two bloqs to perform QVR.
The first is \lstinline|QvrZPow|, which accepts an integer $n$ and a precision $\epsilon$,
and implements $\QVR_{n, \epsilon}$
by decomposing into $n$ single qubit $\ZPow$ rotations
applied to the $n$-bit input register, each with accuracy $\frac{\epsilon}{n}$~\cite[Fig. 14]{Cody_Jones_2012}.
Each single qubit $Z$ rotation can be synthesized using any of the techniques in \cref{subsec:rotations:known-angle}.

The second bloq for QVR is \lstinline{QvrPhaseGradient},
which uses a quantum-quantum addition~\cite{Gidney_2018,PRXQuantum.1.020312} into a phase-gradient state~(\cref{eq:phase-gradient-state}) to obtain a phase kickback on the input register.
The constructor \lstinline{QvrPhaseGradient.from_bitsize} accepts an integer $n$ and a precision $\epsilon$ and implements $\QVR_{n, \epsilon}$.

For ease of exposition, we considered a simplified definition of QVR above.
Qualtran also implements bloqs for more general definitions of QVR~\cite{Cody_Jones_2012,Sanders_2020} which have a constant $\gamma$ programmed into them, and phase by a value $e^{2\pi i \gamma x}$.
It also provides \lstinline{PhasingViaCostFunction}, which accepts a bloq computing function $f(x)$ and phases by $e^{2\pi i f(x)}$~\cite{Sanders_2020}.
For example, the \lstinline{HammingWeightPhasing} bloq uses hamming weight as the cost function,
to apply parallel $\ZPow(t)$ gates to $n$ qubits more efficiently~\cite{Gidney_2018}.

\subsection{Unary Iteration and Indexed Operations} \label{sec:unary-iteration}
The unary iteration primitive is a quantum analogue of a classical \verb`for`-loop. 
Specifically, we wish to construct a unitary operation $U$ that acts on a \emph{selection} register of size $S$ and a \emph{target} register of size $T$; and its action is described as 
\begin{equation}\label{eq:unary_iteration}
    U = \sum_{i=0}^{L-1}\ket{i}\bra{i} \otimes V_{i}
\end{equation}
where $L=2^S$ is the number of unitaries $V_i$ to apply on different branches of the superposition. 
This is the equivalent of a classical \verb`for`-loop with a conditional as follows:
\begin{lstlisting}[language=Python, mathescape=true, basicstyle=\footnotesize\tt]
for $i$ in range($L$):
    if $\emph{selection}$ register $S$ stores integer $i$:
        Apply $V_{i}$ on $\emph{target}$ register $T$
\end{lstlisting}
This primitive has been extensively used in the quantum algorithms literature, albeit by different names. 
\citet{Shende2006} introduced this primitive as a \emph{Quantum Multiplexor} and used it to develop a theory of quantum conditionals and to generalize several known constructions from digital logic, such as Shannon decomposition of Boolean functions. 
\citet{Childs2018} introduced this primitive as the \emph{Select(V)} operation and described an implementation via a tree traversal of a balanced binary tree with $L=2^S$ leaf nodes using $1.5 L - 1$ Toffoli gates and $S$ clean ancilla qubits. 
\citet{Babbush2018Encoding} improved this construction to use $L - 1$ Toffoli gates using $S$ clean ancilla qubits and coined the term \emph{Unary Iteration}. 
\citet{Sanders_2020} describes an optimization for iterating over a sparse set of indices with a Toffoli count that scales as $\min(2^{S} - 1, S \times \abs{L})$ where $\abs{L}$ is the number of indices to iterate upon (i.e. the sparsity), $S$ is still the size of the selection register and now $\abs{L}$ can be $\ll 2^{S}$. 
\citet{khattar2024riseconditionallycleanancillae} extends the construction of~\cite{Babbush2018Encoding, Childs2018} to use skew trees instead of balanced binary trees with either $S$ dirty ancilla qubits and cost $1.25L + \mathcal{O}(\sqrt{L})$ Toffoli or $\log_2^*{S}$ clean ancilla with $2.25L$ Toffoli.

\begin{figure}[H]
\begin{subfigure}[t]{0.49\textwidth}
\begin{lstinputlisting}[language=Python, basicstyle=\tiny\tt]{qrom_apply_lth_bloq_1d.py}
\end{lstinputlisting}
\caption{A coherent \texttt{for}-loop, to load a 1D classical dataset in superposition, implemented using the \lstinline{ApplyLthBloq} multiplexer. The example shows how Qualtran enables users to easily implement, test, get resource estimates and draw nice diagrams for a custom QROM bloq.
}
\label{fig:qrom_apply_lth_bloq:code_1d}
\end{subfigure}
\hfill
\begin{subfigure}[t]{0.49\textwidth}
\begin{lstinputlisting}[language=Python, basicstyle=\tiny\tt]{qrom_apply_lth_bloq_2d.py}
\end{lstinputlisting}
\caption{Nested coherent \texttt{for}-loops can be be expressed by passing multi-dimensional NumPy arrays of Bloqs to the \lstinline{ApplyLthBloq} multiplexer. Example shows a custom QROM to load a 2D classical dataset in superposition using two selection registers and one target register.}
\label{fig:qrom_apply_lth_bloq:code_2d}
\end{subfigure}
\begin{subfigure}[]{\textwidth}
\includegraphics[width=1\linewidth]{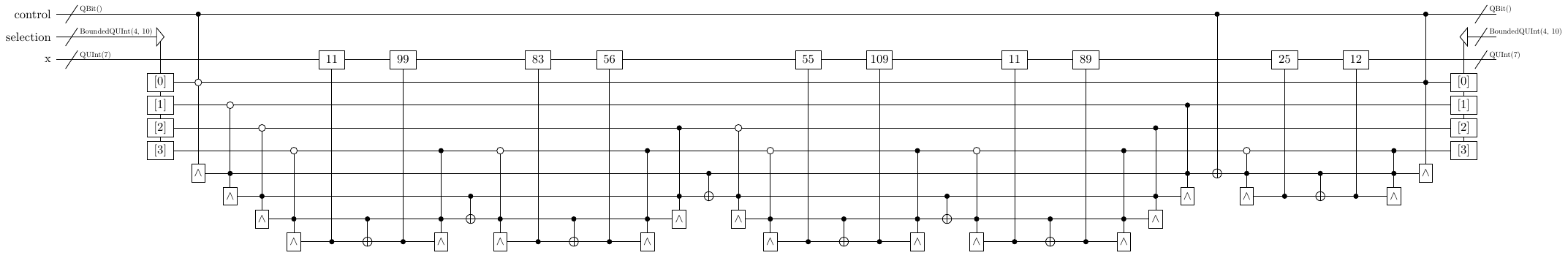}
\caption{Unary iteration circuit to load a list of $N\texttt{=}10$ integers - $[11, 99, 83, 56, 55, 109, 11, 89, 25, 12]$, using \lstinline{ApplyLthBloq} multiplexer. Latex diagram is auto-generated by executing the Qualtran code block in \fig{qrom_apply_lth_bloq:code_1d}.}
\label{fig:qrom_apply_lth_bloq:latex_1d}
\end{subfigure}
\caption{Example of using \lstinline{ApplyLthBloq} multiplexer to execute one ore more nested coherent \texttt{for}-loops. \lstinline{ApplyLthBloq} implements the interface defined by \lstinline{UnaryIterationGate} and uses the unary iteration circuit construction from~\citet{Babbush2018Encoding}. Note that this example is purely for an illustration of how to express coherent for-loops and implement multiplexed operations in Qualtran; for implementing QROMs see \cref{sec:quantum_lookup_table} for a variety of native bloqs that Qualtran provides.}
\end{figure}

In Qualtran, the simplest API to use the unary iteration construction from Ref.~\cite{Babbush2018Encoding} is via the \lstinline{ApplyLthBloq} class. 
Given a list of $\text{bloqs} = [V_{0}, V_{1}, ..., V_{L - 1}]$ that all act on the same target registers, we can construct a bloq \lstinline{U = ApplyLthBloq(bloqs)} such that
$$
    U = \sum_{\ell=0}^{L-1}\ket{\ell}\bra{\ell} \otimes V_{\ell}
$$
The \lstinline{ApplyLthBloq} class also supports specifying an $N$-dimensional NumPy~\cite{harris2020array} array of subbloqs, in which case the bloq allocates $N$ selection registers, one for each dimension, and is equivalent to $N$ nested \verb`for`-loops. 
\fig{qrom_apply_lth_bloq:code_1d} shows how one can use the \lstinline{ApplyLthBloq} to load 1D data in a target register and perform classical simulation to verify the decomposition is correct---i.e. the $i$-th data element is loaded in the target register when the selection register stores integer $i$. \fig{qrom_apply_lth_bloq:code_2d} shows how the construction can be modified to load 2D data using two selection registers.

The \lstinline{ApplyLthBloq} implements the interface defined by the abstract base class \lstinline{UnaryIterationGate}, which can be used directly by power users who need more control on the tree iteration strategy when constructing the unary iteration circuit. 
For example, one can implement the ``variable-spaced QROM" optimization presented in~\cite{Sanders_2020} by overriding the \lstinline{.break_early} method to return $\emph{True}$ whenever all elements in the subtree are identical. Note that this example is purely for illustration and Qualtran provides native bloqs for loading classical data via QROM as described in \cref{sec:quantum_lookup_table}

\subsection{Quantum Lookup Table: Encoding Classical Data as Basis States}\label{sec:quantum_lookup_table}
Similar to a classical lookup table that returns classical data $x_i$ for a queried index $i$, a quantum lookup table can be defined as a unitary $O_{x}$ that responds with a superposition of classical data when queried with an arbitrary superposition of address bits $\sum_{i}\alpha_{i}\ket{i}$
\begin{equation}\label{eq:qrom_oracle}
    O_{x} \sum_{i=0}^{N - 1}\alpha_{i}\ket{i}\ket{0} \rightarrow \sum_{i=0}^{N-1}\alpha_{i}\ket{i}\ket{x_{i}}
\end{equation}

A variety of implementations~\cite{Giovannetti2008, Babbush2018Encoding, Low2024} have been proposed in the literature for the quantum lookup table oracle defined in \eq{qrom_oracle}. We refer the readers to Table~I of \citet{zhu2024unifiedarchitecturequantumlookup} for tradeoffs between the different proposals. 
In Qualtran, we provide built in Bloqs for some of the most commonly used implementations of a Quantum Lookup Table:

\begin{table}[H]
\begin{ruledtabular}
\begin{tabular}{lcccc}
Bloq & Reference & Toffoli & Cliffords & Ancilla \\ 
\hline
\lstinline`QROM` & \citet{Babbush2018Encoding} & 
$N - 2$ & 
$N\cdot b$ & 
$\ceil{\log_2{N}}$ clean\\
\lstinline`SelectSwapQROM` & 
\citet{Low2024} & 
$2\ceil{N / 2^{k}} + 4 b (2^{k} - 1) $ & 
$2 N b + \mathcal{O}(\sqrt{N})$ & 
$2^{k} b$ dirty, $\ceil{\log_2({N/2^{k}})}$ clean \\
\lstinline`QROAMClean` & 
\citet{berry2019qubitization} & 
$\ceil{N / 2^{k}} + b (2^{k} - 1) $ & 
$N b + \mathcal{O}(\sqrt{N})$ & 
$(2^{k} - 1) b + \ceil{\log_2({N/2^{k}})}$ clean \\ 
\lstinline`QROAMCleanAdjoint` &
\citet{berry2019qubitization} & 
$\ceil{N / 2^{k}} + (2^{k} - 1) $ & 
$N + \mathcal{O}(\sqrt{N})$ & 
$2^{k} + \ceil{\log_2({N/2^{k}})}$ clean \\ 
\end{tabular}%
\end{ruledtabular}%
\caption{Bloqs in Qualtran, along with their gate costs and ancilla requirements, that implement a Quantum Lookup Table operation. 
Here $k$ is a configurable parameter that can be used to tradeoff between number of ancilla qubits and the Toffoli gate cost.
All shown bloqs implement a common interface defined by the \lstinline`QROMBase` class and can be substituted for one another to explore different tradeoffs when analyzing a quantum algorithm. 
All shown bloqs support loading \emph{multiple multi-dimensional} classical datasets. Here \emph{multiple} classical datasets imply multiple target registers and \emph{multi-dimensional} classical datasets imply multiple selection registers. 
All shown bloqs also support protocols like symbolic cost analysis, circuit decomposition and classical and quantum simulation to verify correctness. 
} 
\label{tab:qroms_table}
\end{table}

The \lstinline{QROM} bloq implements the construction from Section III.C of Ref.~\cite{Babbush2018Encoding}, which is based on using the unary iteration construction described in \eq{unary_iteration}, and also implements the \emph{variable-spaced QROM} optimization from Figure~3 of Ref.~\cite{Sanders_2020} which reduces gate counts when the data to be loaded is sparse in terms of number of unique elements. 

The \lstinline{SelectSwapQROM} bloq implements the construction from Figure~1d of~\citet{Low2024}, which uses $\mathcal{O}(\sqrt{N})$ additional ancillas to reduce the T / Toffoli gate count by a factor of $~\mathcal{O}(\sqrt{N})$. In this construction, a configurable parameter $K=2^{k}$ is chosen and $b\cdot K$ \emph{dirty} ancilla qubits are used as targets to load a $N/K$ batches of data, each of size $K$. The Toffoli cost of the resulting construction is $2\ceil{N/K} + 4b(K-1)$.

The \lstinline{QROAMClean} bloq implements the optimized variant of \lstinline{SelectSwapQROM} described in Appendix~ A,~B of \citet{berry2019qubitization}, which assumes the target register is always initialized in the $\ket{0}$ state. In this construction, $b \cdot (K-1)$ \emph{clean} ancilla qubits are used as targets to load the batched data, and measurement based uncomputation is used to clean up the clean ancilla registers. The Toffoli cost of the resulting construction $\ceil{N/K} + b(K-1)$.

The \lstinline{QROAMCleanAdjoint} bloq implements measurement based uncomputation of a table lookup via the construction described Appendix~C 
of \citet{berry2019qubitization}, which assumes the target register should be left in the $\ket{0}$ state after uncomputation. This construction reduces a table lookup over $N$ elements, each of target bitsize $b$ to a table lookup of $N/2$ elements, each of target bitsize $2$, by first measuring the target register in the X-basis and using the measurement outcomes to do a corrective table lookup. The corrective table lookup is applied using a construction similar to \lstinline{QROAMClean}. Thus, the Toffoli cost of the resulting construction is $\ceil{N/K} + K$.

All of the bloqs for quantum table lookup described above support loading multiple classical datasets, where each dataset can be multi-dimensional. For \lstinline{SelectSwapQROM} and its variants, the bloq implementation incorporates optimizations to choose multiple block sizes, one for each dimension of the classical dataset, as described in Appendix~G of \citet{Lee2021Even}

\begin{figure}[H]
\begin{adjustwidth}{-2cm}{-2cm}
\begin{subfigure}[c]{0.3\linewidth}
\includegraphics[width=1\linewidth]{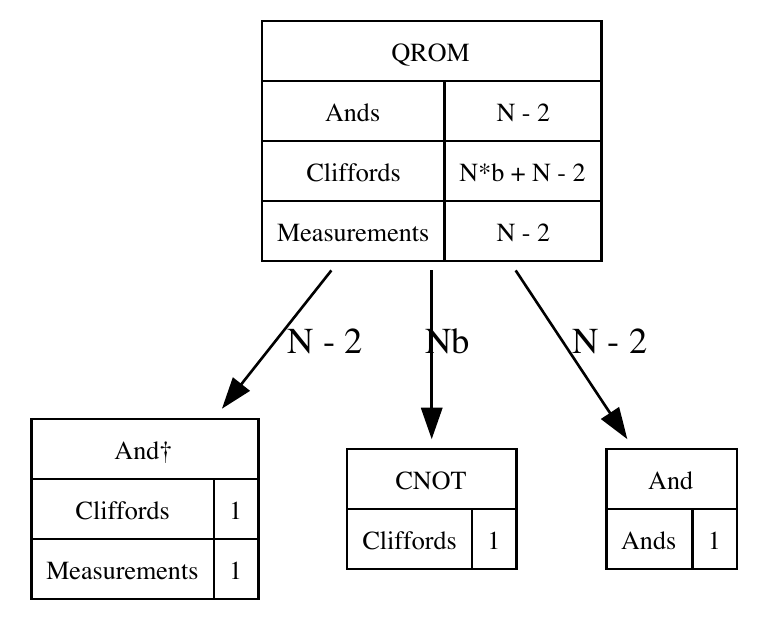}
\caption{Symbolic call graph for \lstinline`QROM` bloq, which implements the construction from \citet{Babbush2018Encoding} and incorporates variable spaced QROM optimization from \citet{Sanders_2020}.}
\end{subfigure}~
\begin{subfigure}[c]{0.5\linewidth}
\includegraphics[width=1\linewidth]{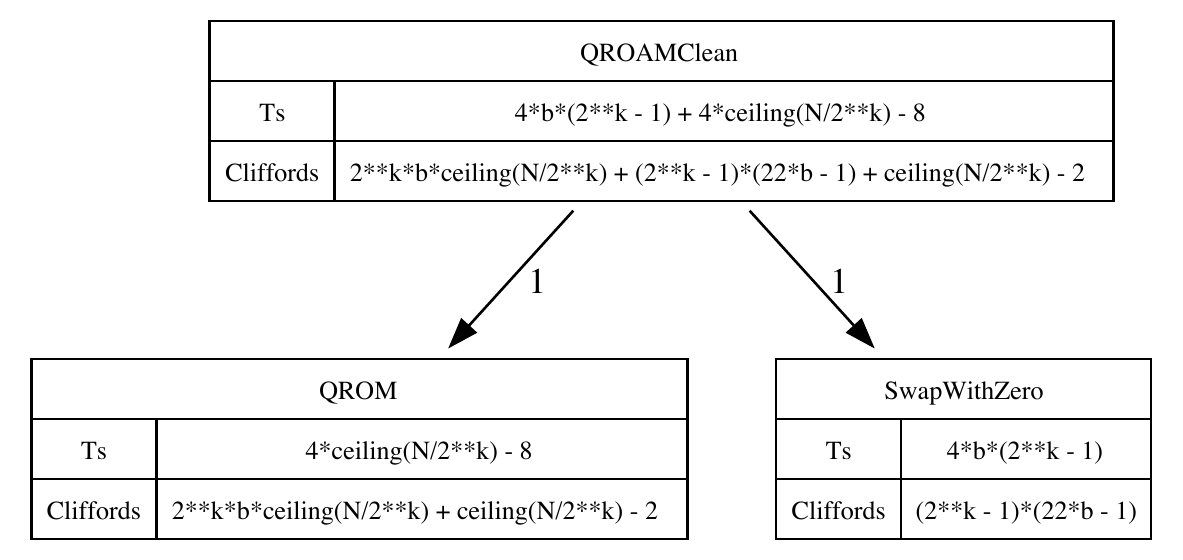}
\caption{Symbolic call graph for \lstinline`QROAMClean` bloq, which implements the clean ancilla construction from Appendix B of \citet{berry2019qubitization}. Decomposes into a single call of \lstinline`QROM` and \lstinline`SwapWithZero` bloqs. Supports loading multi-dimensional data without constructing a contiguous register following optimization from Appendix G of \citet{Lee2021Even}.}
\end{subfigure}
\begin{subfigure}[c]{1\linewidth}
\includegraphics[width=1\linewidth]{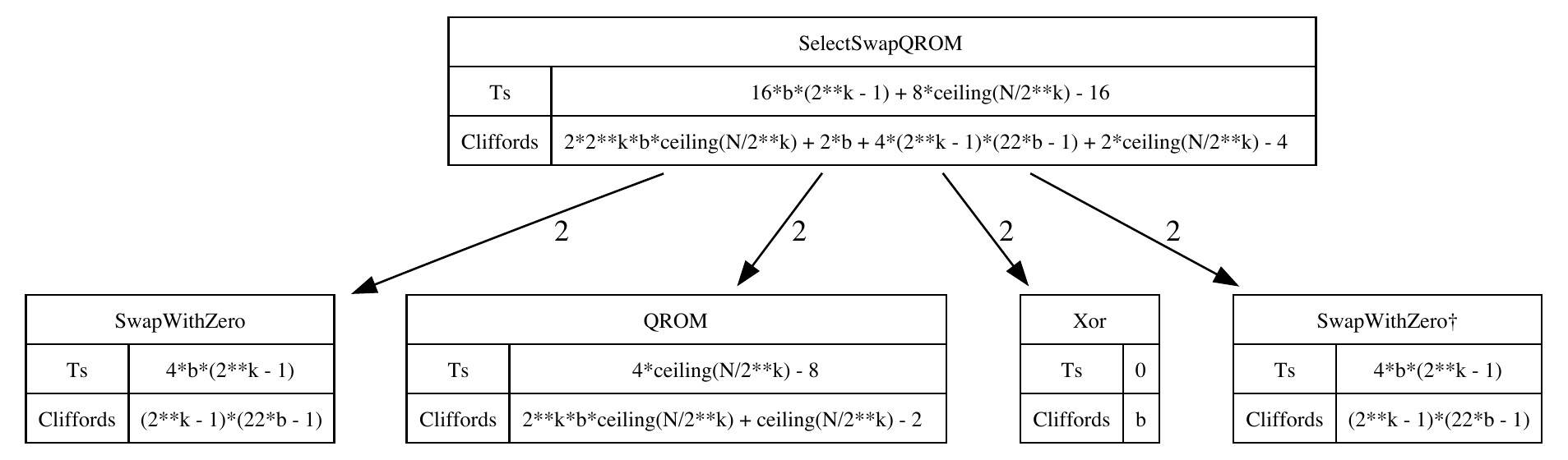}
\caption{Symbolic call graph for \lstinline`SelectSwapQROM` bloq, which implements the original SELECT-SWAP network using dirty ancilla proposed in Figure 1D of \citet{Low2024}. Decomposes into two calls of \lstinline`QROM` and four calls of  \lstinline`SwapWithZero` bloqs. Supports loading multi-dimensional data without constructing a contiguous register following optimization from Appendix G of \citet{Lee2021Even}.}
\end{subfigure}
\centering
\begin{subfigure}[b]{1\linewidth}
\includegraphics[width=1\linewidth]{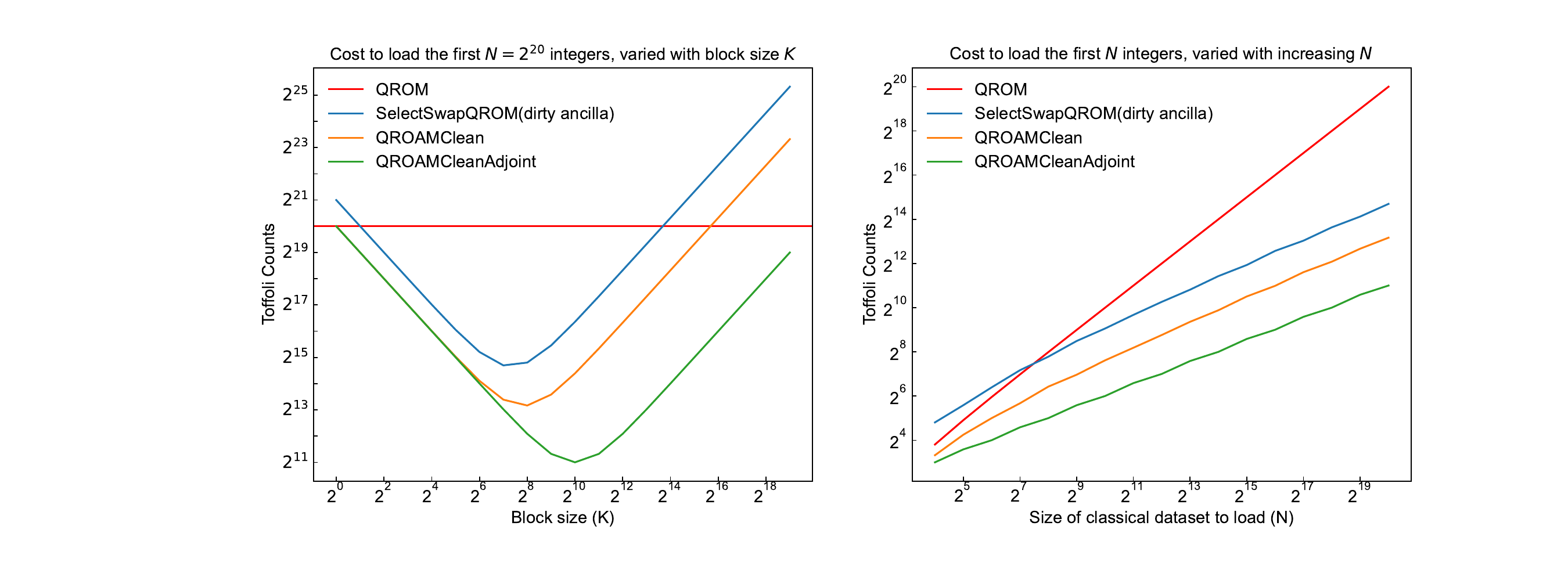}
\caption{Numerical comparison of Toffoli gate counts for bloqs implementing a Quantum Lookup Table to load a 1D classical dataset comprising of the first $N$ integers. Here $k$ is a configurable parameter used to tradeoff Toffoli gates using additional ancilla qubits in bloqs implementing a variation of the SELECT-SWAP network. In the figure on the right, the block size $K$ is chosen optimally to minimize the Toffoli counts.}
\label{fig:qrom_costs}    
\end{subfigure}
\caption{Examples of using Qualtran to analyze the cost of Quantum Lookup Tables. }
\end{adjustwidth}
\end{figure}

\subsection{Quantum State Preparation: Encoding Classical Data as Amplitudes} \label{sec:quantum_state_prep}
Quantum State Preparation is the task to encode an $N$-dimensional vector of classical data 
$\Vec{a} = (a_0, a_1, \dots, a_{N - 1})$ as amplitudes of $N$ computational basis states over $n=\log{N}$ qubits. The problem has been 
widely studied in the quantum algorithms literature~\cite{grover2002creatingsuperpositionscorrespondefficiently, mottonen2004transformationquantumstatesusing, Shende2006, aaronson2016complexityquantumstatestransformations, Low2024, lemieux2024quantumsamplingalgorithmsquantum} 
and has applications across quantum chemistry~\cite{Babbush2018Encoding, berry2019qubitization, low2019hamiltonian, reiher2017elucidating, childs2012hamiltonian}, quantum dynamics~\cite{rubin2024quantum, agrawal2024quantifying}, and solving systems of linear equations~\cite{Harrow_2009,PRXQuantum.3.040303}.

In this section, we describe two variants of the quantum state preparation problem, both of which assume that
classical data to be encoded is given as entries of a classical database and no specific structure about the 
data is assumed. 
For both the variants described below, a sparse version of the problem can be defined analogously where 
the number of non-zero entries in the classical description of the state is specified by a sparsity parameter $S \ll N$.    
Note that when specific structure about the state to be prepared is known, one can often~\cite{huggins2024efficientstatepreparationquantum,lemieux2024quantumsamplingalgorithmsquantum} do better than the arbitrary state 
preparation routines discussed below. 
For example, \lstinline{PrepareUniformSuperposition} bloq in Qualtran can be used to prepare a uniform superposition over 
the first $N$ basis states, where $N$ may not be a power of 2, using a single round of amplitude amplification as described 
in~\citet{Babbush2018Encoding}. 

\subsubsection{Preparation of Arbitrary Quantum States}
The arbitrary quantum state preparation problem seeks to encode classical data described by an $N$ dimensional vector 
of complex numbers $\Vec{a} = (a_0, a_1, \dots, a_{N - 1})$, such that $a_x = \abs{a_x}e^{i\phi(x)}$, 
as amplitudes of a quantum state with $N$ basis vectors prepared on $n=\log_2{N}$ qubits such that the 
prepared state  $\ket{\psi}_\text{prep}$ is $\epsilon$-close to the desired state $\ket{\psi}_f$ defined as

\begin{equation}
    \ket{\psi}_f = \frac{1}{\norm{\Vec{a}}_2}\sum_{x=0}^{N-1}a_x\ket{x}
\end{equation}
    
where $\norm{\Vec{a}}_2 = \sqrt{\sum_{x}\abs{a_x}^2}$ is a normalization constant and 
$$
    \norm{\ket{\psi}_\text{prep} - \ket{\psi}_{f}}_2 \leq \epsilon
$$
where $\norm{.}_2$ is the Euclidean norm.

In Qualtran, the \lstinline{StatePreparationViaRotations} bloq implements the technique for arbitrary state preparation described in~\citet{Low2024}, which reduces state preparation over $n$ qubits and $N=2^{n}$ elements to $n$ data lookups of length $[1, 2, 4, \dots, 2^{n-1}]$ elements, each loading a $b$-bit approximation of a rotation angle. Each data lookup is followed by $b$ single-qubit controlled rotations. 
Thus, a total of $N=1 + 2 + \dots + 2^{n - 1}$ data elements are loaded, each of bitsize $b$, across $n$ different lookup calls and $\mathcal{O}(n \times b)$ single qubit controlled rotations are applied. 
Each data lookup step is implemented via techniques described in \sect{quantum_lookup_table} and the single qubit rotations are synthesized via techniques described in \sect{rotations}. Here $b=\log_2({n/\epsilon})$ is chosen such that the prepared state is $\epsilon$-close to the desired state. 
The $n$ data lookups to load $N=2^n$ rotation angles, each of size $b=\log_2({n/\epsilon})$,  dominate the overall cost of the algorithm.

For sparse state preparation, \lstinline{SparseStatePreparationViaRotations} can be used which reduces sparse state preparation to dense state preparation
via techniques described in~\citet{ramacciotti2023simplequantumalgorithmefficiently}. 

\subsubsection{Preparation of Arbitrary Purified Density Matrices}
Arbitrary purified density matrix preparation seeks to encode classical data described by an $N$-dimensional vector of positive real numbers $\Vec{w} = (w_0, w_1, \dots, w_{N - 1})$, 
such that $w_i \ge 0 \land w_i \in \mathcal{R}$, as probabilities of an ensemble of $N$ basis states represented by a density matrix $\rho_\text{prep}$ which is $\epsilon$-close to the desired density matrix  $\rho_f$ defined as 
\begin{equation}
\rho_f = \sum_{x=0}^{N - 1}\frac{w_{x}}{\norm{\Vec{w}}_1}\ket{x} \bra{x} 
\end{equation}

where $\norm{\Vec{w}}_1 = \sum_{x}w_x$ is a normalization constant and 
$$
    \norm{\rho_\text{prep} - \rho_f}_{1} \leq \epsilon    
$$
where $\norm{.}_{1}$ is the trace distance. 
This is achieved by preparing a purification of $\rho_\text{prep}$ of the form
\begin{equation}
    \ket{\psi}_\text{prep} = \sqrt{\frac{\widetilde{w_{x}}}{\norm{w}_1}}\ket{x} \ket{\text{temp}_{x}}
\end{equation}
where the desired number state $\ket{x}$ is allowed to be entangled with a garbage state $\ket{\text{temp}_{x}}$ that depends only on $x$. Such a state is often cheaper to prepare than preparation of arbitrary quantum states described in the previous section, and is sufficient for applications like quantum simulation based on linear combination of unitaries or qubitization \cite{Babbush2018Encoding, low2019hamiltonian, Berry2015}.

In Qualtran, the \lstinline{StatePreparationAliasSampling} bloq implements the technique for arbitrary purified density matrix preparation described in \citet{Babbush2018Encoding}, which reduces the problem of state preparation over $n + \mu$ qubits and $N=2^{n}$ elements to that of a single data lookup to load $N$ elements, each of bitsize $n + \mu$, followed by a quantum-quantum comparison and conditional swap. 
Here $\mu=\lceil\log_2{1/\epsilon}\rceil$ is chosen such that the prepared state is $\epsilon$-close to the desired state.
The data lookup step dominates the overall cost and is implemented via techniques described in \sect{quantum_lookup_table}

For sparse state preparation, \lstinline{SparseStatePreparationAliasSampling} can be used which reduces sparse state preparation to dense state preparation
via techniques described in Section~5 of \citet{berry2019qubitization}.

\subsection{Block Encoding}
\label{sec:block-encoding}
Numerous quantum algorithms, including those for Hamiltonian simulation~\cite{Lee2021Even, vonburg2021catalysis}, linear algebra~\cite{Harrow_2009, chakraborty2019}, and portfolio optimization~\cite{Dalzell_2023}, rely on loading a matrix of data or transforming a quantum state via a matrix. To do so, they utilize \emph{block encoding} --- a way to represent a matrix as a unitary operator when the matrix is not unitary, or is inconvenient or inefficient to directly realize as a unitary operator.

Let $A$ be an $s$-qubit linear operator, which is not necessarily unitary. Following Ref.~\cite{gilyen2019quantum}, we say that the $(s + a)$-qubit unitary $\mathcal{B}[A]$ is a $(\alpha, a, \epsilon)$-block encoding of $A$ if:
$$\norm{A - \alpha(\bra{0}^{\otimes a} \otimes I) \mathcal{B}[A] (\ket{0}^{\otimes a} \otimes I)} \le \epsilon$$
where $\alpha$ is a normalization constant chosen such that $\norm{A}/\alpha \le 1$, and $\epsilon \ge 0$ is the precision of the encoding. In other words, the matrix $\mathcal{B}[A]$ contains $A/\alpha$ in its upper-left block:
$$
\mathcal{B}[A] = \begin{pmatrix}
A/\alpha & * \\
* & *
\end{pmatrix}
$$
Historically, the concept of block encoding emerged from methods to encode matrices as the input to algorithms for quantum linear algebra~\cite{Harrow_2009} and from methods to encode Hamiltonians as a linear combination of unitaries~\cite{childs2012hamiltonian}. It gained prominence with the development of quantum recommendation systems~\cite{kerenidis2016quantumrecommendationsystems} and optimal Hamiltonian simulation by qubitization~\cite{low2019hamiltonian}. It cemented its role at the forefront of interest in algorithms research with the development of the quantum singular value transformation~\cite{gilyen2019quantum}.

Qualtran implements techniques to block-encode various matrices important to quantum algorithms.
First, it provides bloqs to construct block encodings from primitive elements such as unitaries, LCU SELECT/PREPARE oracles, and sparse access matrices. Next, it provides bloqs to compose these block encodings by arithmetic operations such as products and linear combinations of the encoded matrices.

\subsubsection{Primitive Block Encodings from Unitaries and Oracles}

First, in the simplest case, any unitary operator $U$ is a $(1, 0, 0)$-block encoding of $U$. Correspondingly, the \lstinline`Unitary` bloq in Qualtran wraps a unitary bloq into a block encoding with $\alpha = 1$.

Second, any linear combination of unitaries $A = \sum_i \lambda_i A_i$ as specified by a pair of SELECT and PREPARE oracles defines a block encoding of $A$, which can be created by Qualtran's \lstinline`LCUBlockEncoding` bloq.

Third, Qualtran's \lstinline`SparseMatrix` bloq constructs a block encoding for a sparse matrix $A$ with at most $m$ non-zero entries in each row or column, given oracles
\begin{align*}
O_r \ket{i}\ket{k} = \ket{i}\ket{r_{ik}} \\
O_c \ket{j}\ket{\ell} = \ket{c_{j\ell}}\ket{\ell}
\end{align*}
to compute the index $r_{ij}$ or $c_{ji}$ of the $j$-th nonzero entry in the $i$-th row or column respectively of $A$, and
$$
O_A \ket{i}\ket{j}\ket{0} = \ket{i}\ket{j}\ket{A_{ij}}
$$
to compute the matrix entry $A_{ij}$~\cite{Dalzell_2023}.

Qualtran provides a set of pre-implemented row/column oracles for matrices where the non-zero elements are located in the top-left block (\lstinline`TopLeftRowColumnOracle`) or within a band along the main diagonal (\lstinline`SymmetricBandedRowColumnOracle`). It also provides pre-implemented entry oracles for matrices where all elements have equal value (\lstinline`UniformEntryOracle`) or are explicitly given by a NumPy array that is loaded from QROM (\lstinline`ExplicitEntryOracle`). In the more general case, Qualtran gives interfaces and subroutines to aid the user in implementing a custom oracle.

Dense matrices can be encoded as a special case of sparse matrices. In \cref{fig:block_encoding:sparse_matrix}, we present an example of the encoding of a dense $2 \times 2$ matrix explicitly given as a NumPy array. The example uses \lstinline`TopLeftRowColumnOracle` to specify that the top left $2^1 \times 2^1$ block, i.e.\ the entire matrix, is nonzero. It uses \lstinline`ExplicitEntryOracle` to specify the entries to be loaded from QROM. Though this example block encoding incurs larger costs than a sparse matrix given by a more efficient entry access oracle, it shows how a developer can quickly prototype the loading of a dataset or transformation in an algorithm.

\subsubsection{Composite Block Encodings from Matrix Arithmetic}

\begin{table*}
\begin{ruledtabular}
\begin{tabular}{lcccc}
Bloq & block-encodes & normalization $\alpha$ & \# ancillas $a$ & precision $\epsilon$ \\ 
\hline
\lstinline`Unitary` & unitary $U$ & 1 & 0 & 0 \\
\lstinline`SparseMatrix` & $m$-sparse $A$ & $m$ & $\dim(A)+1$ & user-specified \\
\lstinline`TensorProduct` & $\bigotimes_i A_i$ & $\prod_i \alpha_i$ &  $\sum_i a_i$ & $\sum_i \alpha_i \epsilon_i$ \\ 
\lstinline`Product` & $\prod_i A_i$ & $\prod_i \alpha_i$ & $n - 1 + \max_i a_i$ & $\sum_i \alpha_i \epsilon_i$ \\ 
\lstinline`Phase` & $\exp{i\phi}A_1$ & $\alpha_1$ & $a_1$ & $\epsilon_1$ \\ 
\lstinline`LinearCombination` & $\sum_i \lambda_i A_i$ & $\sum_i \lvert\lambda_i\rvert\alpha_i$ & $\lceil \log_2 n \rceil + \max_i a_i$ & $(\sum_i \lvert\lambda_i\rvert)\max_i \epsilon_i$ \\ 
\lstinline`ChebyshevPolynomial` & $T_j(A_1)$ & 1, given $\alpha_1 = 1$ & $a_1$ & $\epsilon_1$
\end{tabular}%
\end{ruledtabular}%
\caption{Selection of block encodings implemented in Qualtran alongside their normalization factor $\alpha$, ancilla overhead $a$, and precision $\epsilon$. Rather than manually reasoning about these performance parameters, users can compute them automatically by invoking the bloq implementations.} \label{tab:block-encodings}
\end{table*}
The full power of block encodings extends not merely to constructing a matrix within a quantum algorithm but also to performing a broad range of arithmetic operations, such as products and linear combinations, on the encoded matrices. Such arithmetic can be used, for example, to construct a Taylor series of $1/A$ in a quantum linear solver~\cite{Harrow_2009} or a Jacobi-Anger expansion of $\sin(At)$ in Hamiltonian simulation~\cite{low2019hamiltonian}.

Qualtran provides a library of composite block encodings to perform arithmetic operations over the primitive forms of block encoding above; we summarize them in \Cref{tab:block-encodings}.
Given a sequence of block encodings $\{\mathcal{B}[A_i]\}_i$, Qualtran provides the \lstinline`TensorProduct` bloq to compute $\mathcal{B}[\bigotimes_i A_i]$ and \lstinline`Product` to compute $\mathcal{B}[\prod_i A_i]$. It provides the bloqs \lstinline`Phase` to compute $\mathcal{B}[\exp{i\phi}A_1]$ and \lstinline`LinearCombination` to compute $\mathcal{B}[\sum_i \lambda_i A_i]$ given appropriate $0 \le \phi < 2\pi$ or $\lambda_i \in \mathbb{R}$ respectively~\cite{gilyen2019quantum}.

In \cref{fig:block_encoding:linear_combination}, we illustrate how a developer uses Qualtran's block encoding library to perform arithmetic on the encoded matrices. Specifically, the example constructs a block encoding of the Chebyshev polynomial $T_2(A) = 2A^2 - I$ given a block encoding of $A$. It does so by invoking the \lstinline`Unitary` bloq to block-encode $I$, invoking the \lstinline`Product` bloq on $A$ with itself, and then invoking the \lstinline`LinearCombination` bloq to produce the final polynomial. In \cref{fig:block_encoding:diagram}, we visualize the more complex underlying sequence of bloq operations to which the high-level programs in \cref{fig:block_encoding:sparse_matrix,fig:block_encoding:linear_combination} compile.

\subsubsection{Optimizing and Analyzing Block Encodings}

\begin{figure*}
\begin{subfigure}[t]{0.45\textwidth}
\begin{lstinputlisting}[language=Python, basicstyle=\tiny\tt]{explicit_matrix.py}
\end{lstinputlisting}
\caption{Explicitly specified matrix $A = \begin{bmatrix}
    0 & 1/4 \\ 1/3 & 0.467
\end{bmatrix}$.}
\label{fig:block_encoding:sparse_matrix}
\end{subfigure}
\hfill
\begin{subfigure}[t]{0.45\textwidth}
\begin{lstinputlisting}[language=Python, basicstyle=\tiny\tt]{linear_combination.py}
\end{lstinputlisting}
\caption{Chebyshev polynomial $T_2(A) = 2A^2 - I$.}
\label{fig:block_encoding:linear_combination}
\end{subfigure}
\begin{subfigure}[]{\textwidth}
\includegraphics[width=1\linewidth]{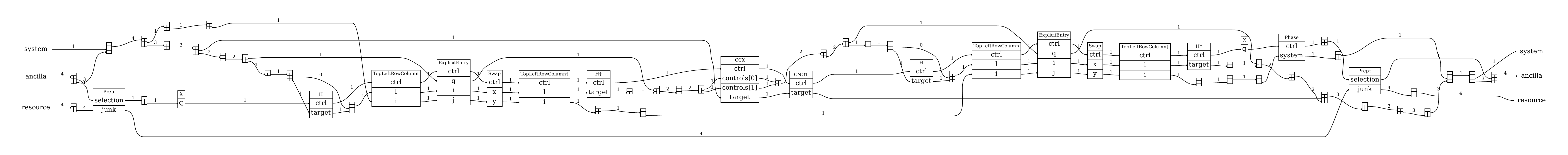}
\caption{Bloq decomposition of the combined programs in \Cref{fig:block_encoding:sparse_matrix,fig:block_encoding:linear_combination}. This diagram is intentionally presented in its full complexity to contrast with the simplicity of the user-facing abstractions above.}
\label{fig:block_encoding:diagram}
\end{subfigure}
\caption{Examples of using Qualtran to construct and transform block encodings of matrices.}
\end{figure*}

Qualtran also provides more optimized implementations for certain forms of arithmetic with appropriate structure. One example is the \lstinline`ChebyshevPolynomial` bloq, described in \cref{subsec:qubitization}, that can implement the transformation of \cref{fig:block_encoding:linear_combination} using fewer ancillas and gates. The user can further improve the efficiency of arithmetic by rewriting its algebraic structure. For example, Qualtran provides a function \lstinline{collect_like_terms} to flatten nested linear combinations and thereby save qubits and gates.

To the researcher, one useful feature is that Qualtran can automatically analyze the properties---normalization factor $\alpha$, ancilla overhead $a$, and precision $\epsilon$---that determine the efficiency of a block encoding and in turn the resource consumption of the quantum algorithm.
Under the hood, Qualtran leverages its support for symbolics to automatically derive values for these parameters from the constituents of a composite block encoding, e.g. the terms being added in a \lstinline`LinearCombination`.

\subsection{Quantum Signal Processing}
\label{sec:qsp}

Quantum Signal Processing (QSP) is a quantum algorithmic primitive to which many standard quantum algorithms can be reduced.
Given a unitary $U$ and a polynomial $P$, QSP implements a block-encoding~(\cref{sec:block-encoding}) of $P(U)$.
QSP is a building block for Quantum Singular Value Transformation (QSVT)~\cite{gilyen2019quantum},
which transforms singular values of a block-encoded matrix.
QSVT has been used to design algorithms for many problems, including quantum linear systems~\cite{chakraborty2019} and Hamiltonian simulation~\cite{low2019hamiltonian}.
See Refs.~\cite{grandunification2021,gilyen2019quantum} for an exposition.

Qualtran provides a \lstinline|GeneralizedQSP| bloq~\cite{motlagh2024generalized}
which allows applying a complex polynomial $P$
to obtain a block-encoding of $P(U)$.
The bloq is parameterized by a pair of complex polynomials $P, Q$, and a unitary bloq $U$,
and implements the unitary
\begin{equation*}
  \mathsf{GQSP}_{P, Q}(U) = \begin{bmatrix} P(U) & \cdot~ \\ Q(U) & \cdot~ \end{bmatrix}
\end{equation*}
The polynomial $P$ above should satisfy $\abs{P(e^{i\theta})} \le 1$ for all $\theta \in [0, 2\pi)$.
For a degree $d$ polynomial $P$,
the bloq uses $d$ calls to controlled-$U$,
along with $d + 1$ arbitrary single qubit rotations.

While the polynomial transformation $P$ is usually defined by the algorithmic designer, it can be challenging to compute the complementary polynomial $Q$. $Q$ should be constructed such that 
\[
\abs{P(e^{i\theta})}^2 + \abs{Q(e^{i\theta})}^2 = 1
\quad\quad
\text{for all $\theta \in [0, 2\pi)$.}
\]

Qualtran implements three methods to compute the complementary polynomial $Q$ provided an input $P$:
\begin{enumerate}
\item \lstinline{qsp_complementary_polynomial}: By factorizing $P$~\cite[Theorem 4]{motlagh2024generalized} (default).
\item \lstinline{fast_complementary_polynomial}: Convex optimization~\cite[Theorem 4]{motlagh2024generalized}.
\item \lstinline{fft_complementary_polynomial}: A fast Fourier transform-based method~\cite[Algorithm 2]{berntson2024complementarygqsp}.
\end{enumerate}

\begin{figure}[H]
 \begin{subfigure}{\textwidth}
   \includegraphics[width=\textwidth]{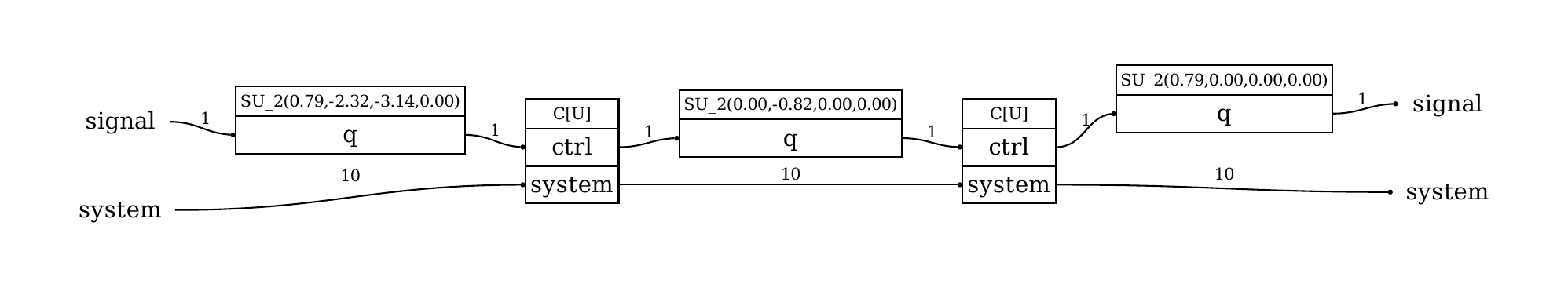}
 \end{subfigure}
   \caption{Generalized QSP for a degree 2 polynomial $P(x) = (x^2 + 1)/2$. The
   register named \emph{signal} is a single qubit on which QSP rotations are
   performed, and $U$ acts on the ten qubit \emph{system} register.
   Each  SU\_$2$ rotation is parameterized by three Euler angles. 
   \label{fig:gqsp}
   }
 \end{figure}

\Cref{fig:gqsp} shows an example GQSP circuit implementing the polynomial $P(x) = (x^2 + 1)/2$ using Algorithm 1 of Ref.~\cite{motlagh2024generalized}.

\subsection{Quantum Phase Estimation} \label{sec:phase-estimation}
In this section, we detail Qualtran's support for a variety of phase estimation bloqs, primarily distinguished by their use of different window functions to minimize phase measurement uncertainty - quantified by metrics like probability of error outside a confidence interval and Holevo variance.  We begin with a concise mathematical overview essential for describing the implemented bloqs, followed by their applications in quantum metrology and simulation. A summary of supported bloqs can be found in \cref{tbl:qpe_costs}.

Given a unitary $U$ with an eigenvector $\ket{\psi}$ and eigenvalue $e^{2\pi i \phi}$, such that $U\ket{\psi} = e^{2\pi i \phi}\ket{\psi}$, Quantum Phase Estimation (QPE) aims to output an estimate of $\phi$ - the eigenphase corresponding to $\ket{\psi}$~\cite{kitaev1995quantummeasurementsabelianstabilizer, Abrams_1999, nielsenchuang}. 
When (a) the eigenstate state $\ket{\psi}$ is exactly prepared and given as an input and (b) the eigenphase $\phi$ is a dyadic rational (i.e. it can be represented exactly with a finite number of bits - $m$), then the action of $\text{QPE}$ can be described as
$$
    \text{QPE}\ket{\psi}\ket{0}^{m} \rightarrow \ket{\psi}\ket{\phi}
$$
In practice, however, both the assumptions (a) and (b) stated above are hard to achieve. In a more general setting, given an input unitary $U$ with an eigendecomposition $U=\sum_{j}e^{2\pi i \phi_j}\ket{\psi_j}\bra{\psi_j}$, an input initial state $\ket{\Psi}=\sum_{j}\gamma_{j}\ket{\psi_j}$ and an auxiliary control register $\ket{0}^m$, QPE allows us to sample from the eigenspectrum of $U$ by applying the map
\begin{equation}\label{eq:qpe_action}
    \text{QPE}\ket{\Psi}\ket{0}^{m} \rightarrow 
    \sum_{j}\sum_{k}\gamma_{j}\tau(j, k)\ket{\psi_j}\tilde{\ket{\phi_k}}
\end{equation}
where $\tau(x, y)$ is a kernel function and $\tilde{\phi_j}$ is an $m$-bit approximation to the desired phase $\phi_j$. If $\forall_j$, $\phi_j$ can be represented exactly using $m$ bits then $\tau(x, y) = \delta_{x, y}$
such that $\text{QPE}\ket{\Psi}\ket{0}^{m} \rightarrow \sum_{j}\gamma_{j}\ket{\psi_j}\ket{\phi_j}$.           

From \Cref{eq:qpe_action}, one can see that on measuring the $m$-bit auxiliary control register in the computational basis, the probability of getting a phase $\tilde{\ket{\phi_k}}$ can be written as
\begin{equation}
    \Pr[\hat{\phi} = \tilde{\phi_{k}}] = \sum_{j} \abs{\gamma_{j}}^2 \abs{\tau(j, k)}^2
\end{equation}
Here $\hat{\phi}$ is a random variable that corresponds to the phase estimate, and can be decomposed as
\begin{equation}
    \hat{\phi} = \phi + \Delta\phi
\end{equation}
where $\phi$ is a random variable satisfying $\Pr[\phi = \phi_j] = \abs{\gamma_j}^2$ and $\Delta\phi$ is a random
variable satisfying conditional distribution $\Pr[\Delta\phi = \hat{\phi_k} - \phi_k | \phi_k] =\abs{\tau(j, k)}^2$.
Here $\phi$ represents the likelihood of projecting the initial state $\ket{\Psi}$ into the desired eigenstate $\ket{\psi_j}$, which depends solely on the initial overlap $\gamma_{j}$, 
and $\Delta\phi$ represents a \emph{phase error} introduced due to \emph{spectral leakage}~\cite{greenaway2024casestudyqsvtassessment, Xiong_2022}, a phenomenon where probability spreads across multiple peaks due to misalignment between the true phase $\phi_j$ and the discrete values $\tilde{\phi_k}$ used to approximate $\phi_j$ using a finite number of bits $m$. 

Window functions are tools borrowed from classical signal processing that can be used to smooth out the sharp transitions at the edges of a probability distribution and concentrate probability mass around the true phase~\cite{Harris1978}.
In QPE, the state preparation unitary used to construct an initial state on the auxiliary control register of size $m$, is referred to as a \emph{window state}. In Qualtran, \lstinline{QPEWindowStateBase} class defines an interface for bloqs that can be used to prepare window states for QPE.

\lstinline{TextbookQPE} bloq in Qualtran accepts any unitary bloq $U$, an instance of \lstinline{QPEWindowStateBase} for preparing an initial window state on the control register of bitsize $m$ and applies $m$ controlled unitaries, $\text{C-U}^{2^{i}}$ in the $i$'th step, followed by an inverse QFT. 

\lstinline{QubitizationQPE} bloq in Qualtran accepts an instance of  \lstinline{QubitizationWalkOperator} $W$~\cite{low2019hamiltonian}, an instance of
\lstinline{QPEWindowStateBase} for preparing an initial window state on the control register of bitsize $m$, and implements the optimization from Figure 2 of \citet{Babbush2018Encoding} to double the phase difference and apply powers of controlled walk operator by only controlling the reflection operator. 

In Qualtran, for both the QPE bloqs desribed above, the cost of QPE on $m$ control qubits is reported as a sum of costs for
\begin{enumerate}
\item \text{CtrlStatePrep}: Cost to prepare the window state on $m$ control qubits, typically scales as $\mathcal{O}(m)$. For the textbook version of QPE~\cite{nielsenchuang}, the window state is a Rectangular window - prepared by simply applying a Hadamard gate on all $m$ control qubits.
\item \text{Controlled-U}'s: There are two cases:    
    \begin{itemize}
        \item If the unitary is fast forwardable; i.e. cost of $\text{C-U}^n$ scales sublinearly with $n$, the cost of this step is at least $m \times \text{cost(C-U)}$
        \item If the unitary is not fast forwardable; i.e. cost of $\text{C-U}^n$ is $n\times\text{cost(C-U)}$, the cost of this step is $2^{m} \times \text{cost(C-U)}$.
    \end{itemize}
\item  $\text{QFT}^\dagger$: The textbook version of inverse QFT uses $\mathcal{O}(m^2)$ 
rotations~\cite{nielsenchuang} but this can be improved to $\mathcal{O}(m \log{m})$ using 
approximate QFT constructions~\cite{Nam_2020, Barenco_1996}.
\end{enumerate}
From the enumeration above one can see that in general, when the unitary $U$ is not fast forwardable, the number of queries made by the Quantum Phase Estimation (QPE) procedure to the controlled unitary $\text{C-U}$ scales exponentially with the number of control qubits $m$. 
Qualtran provides bloqs to prepare a variety of different initial window states, summarized in \tbl{qpe_costs}, which can be used to determine the minimum of number of control qubits $m$ required to keep the uncertainty in the measured phase below a certain threshold $\epsilon$. 

Below, we describe the two most commonly used measures of uncertainty in the estimated phase and corresponding Qualtran bloqs to prepare window states which are optimal for minimizing the uncertainty. 

\begin{table}[H]
\begin{adjustwidth}{-2cm}{-2cm}
\centering
\renewcommand{\arraystretch}{2}
\begin{tabular}{|p{3.8cm}|p{4.3cm}|p{4.3cm}|p{1.7cm}|p{2.5cm}|p{1.7cm}|}
    \hline
    \multicolumn{2}{|c|}{Window Function} & \multicolumn{2}{c|}{Confidence Interval ($\epsilon, \delta$)} & \multicolumn{2}{c|}{Holevo Variance ($\epsilon^2$)} \\
    \hline
        Bloq Name & State & Bitsize-$m$ & Query Cost & Bitsize-$m$ & Query Cost \\
    \hline
        \lstinline`RectangularWindowState` 
        
        (Rectangular Window~\cite{nielsenchuang})
        & $\sum\limits_{x=0}^{2^m - 1}\frac{1}{\sqrt{2^m}}\ket{x}$ 
        & $\log_2(\frac{1}{\epsilon})+\log_2(2\texttt{+}\frac{2}{\delta})$ 
        
        \cite{nielsenchuang, Cleve1998} 
        & $\mathcal{O}(\frac{1}{\epsilon \delta})$ 
        & $2\log_2(\frac{\pi}{\epsilon})$~\cite{nielsenchuang, Higgins2007}
        & $\mathcal{O}\left(\frac{\pi^2}{\epsilon^2}\right)$   \\

    \hline
        \lstinline`LPResourceState`         
        
        (Sine Window~\cite{Luis1996})
        & $\sum\limits_{x=0}^{2^m - 1}\sin\left(\frac{\pi (x+1)}{2^m + 1}\right)\ket{x}$ 
        & $\log_2(\frac{1}{\epsilon}) + 
        \log_2(\frac{\pi^{2/3}}{48^{1/3}\delta^{1/3}} \texttt{+} 2)$ 
        
        \cite{Rendon_2022}
        & $\mathcal{O}(\frac{1}{\epsilon} \frac{1}{\delta^{1/3}})$ 
        & $\log_2(\frac{\pi}{\epsilon})$ \cite{Babbush2018Encoding}
        & $\mathcal{O}\left(\frac{\pi}{\epsilon}\right)$ \\
    \hline
        \lstinline`KaiserWindowState`
        
        (Kaiser Window~\cite{Kaiser1980})
        & $\sum\limits_{x=-M}^{M}\frac{1}{2M} \frac{I_0\left(\pi\alpha\sqrt{1-(x/M)^2}\right)}{I_0\left(\pi\alpha\right)}\ket{x}$ 
        & $\log_2(\frac{1}{\epsilon}\ln{\frac{1}{\delta}})+
        \mathcal{O}(\log_2{(\frac{\ln\ln{\frac{1}{\delta}}}{\epsilon})})$ 
        
        \cite{Berry_2024, greenaway2024casestudyqsvtassessment}
        & $\mathcal{O}(\frac{1}{\epsilon}\ln{\frac{1}{\delta}})$
        & 
        &   \\
    \hline
\end{tabular}
\caption{Comparison of window functions used in Quantum Phase Estimation (QPE). In Qualtran, \lstinline`TextbookQPE` bloq can be used to perform phase estimation on an input unitary $U$. When the unitary is a qubitized walk operator, one can use \lstinline`QubitizationQPE` blow which incorporates optimizations from \citet{Babbush2018Encoding}. Both the QPE bloqs accept an instance of \lstinline`QPEWindowStateBase` to prepare a window state on the control register. The table shows bloqs implementing three commonly used window states in the literature along with their impact on the two commonly used measures of uncertainty in the estimated phase. \lstinline`LPResourceState` are optimal to minimize the Holevo variance ($V_H = \epsilon^2$) \cite{Luis1996} and \lstinline`KaiserWindowState` are optimal to minimize probability of error ($\delta$) outside a given confidence interval ($\epsilon$) given as $\Pr\left[\Delta\phi > \epsilon\right] \le \delta$. \cite{patel2024optimalcoherentquantumphase}.}
\label{tbl:qpe_costs}
\end{adjustwidth}
\end{table}

\subsubsection{Minimizing Probability of Error Outside a Confidence Interval}
One way to characterize the uncertainty $\Delta \phi$ is to bound the probability with which the measured phase $\tilde{\phi}$ is outside of a confidence interval around the true phase $\phi$ of half-width $\epsilon$, defined as $[\phi - \epsilon, \phi + \epsilon]$. This version is also known as \emph{high-confidence QPE}

$$
    \Pr\left[\Delta\phi > \epsilon\right] \le \delta
$$
It is known that a phase estimation protocol with precision $\epsilon$ and error probability $\delta$ has cost
$\Omega(\frac{1}{\epsilon}\log{\frac{1}{\delta}})$\cite{mande2023tightboundsquantumphase} and this can be achieved by using \emph{Kaiser} window based control states~\cite{Berry_2024, greenaway2024casestudyqsvtassessment} or \emph{Discrete Prolate Spheroidal Sequence (DPSS)}~\cite{SlepianPollak1961, Imai2009, patel2024optimalcoherentquantumphase} based control states. In Qualtran, the \lstinline{KaiserWindowState} bloq can be used to initialize the control register in the \emph{Kaiser} window state, given a specific choice of parameter $\alpha$, and uses bloqs for arbitrary state preparation described in \sect{quantum_state_prep}.

Minimizing the probability of error outside a confidence interval is a useful measure of uncertainty in applications where
\begin{itemize}
    \item The relationship between the desired quantity and estimated phase is nonlinear and the application is highly sensitive to errors outside the confidence interval. For example - amplitude estimation in~\cite{Berry_2024}, variable time amplitude amplification in~\cite{Childs2017} or ground state energy estimation by taking minimum of sampled values. 
    \item When the choice of parameter $\delta$ is well motivated, the high confidence version leads to optimal constant factors.
\end{itemize}
\subsubsection{Minimizing Holevo Variance}
The Holevo variance, defined as $V_H =  \langle \cos{\Delta \phi}\rangle^{-2} - 1$, is a convenient measure of variance for phase because it enables simple analytic results and is close to the mean-square error for narrowly peaked distributions~\cite{Holevo1984, Berry2009}. 
Specifically, it can be shown~\cite{Babbush2018Encoding, najafi2023optimumphaseestimationcontrol} that there exists a phase estimation protocol which uses $m$ control qubits and $2^m$ applications of $\text{Controlled-U}$ to yield an estimate of the phase $\tilde{\phi}$ such that 
$$
    \mathbb{E}(\Delta\phi) = 0 \text{ and } V_H(\Delta \phi) = \tan^2\left({\frac{\pi}{2^m + 1}}\right) \approx \left(\frac{\pi}{2^m}\right)^2
$$

In Qualtran, \lstinline{LPResourceState} bloq can be used to initialize the control register in \emph{Sine} state, which are optimal control states~\cite{Luis1996} to minimize the Holevo variance, using a single round of amplitude amplification via the procedure described in~\cite{Babbush2018Encoding}. 

Holevo variance is a useful measure of uncertainty in applications where
\begin{itemize}
    \item The estimated phase is treated as a continuous quantity and the magnitude of the error is also important. For example in quantum metrology~\cite{Berry2009}.
    \item The relationship between the desired quantity and estimated phase is linear, so one can take advantage of variance bounds on the estimated phase to obtain variance bounds on the desired quantity. For example - finite difference estimation of forces from~\cite{OBrien2022}.
    \item The variance version is easy to get rough resource estimates since its parameter free. The confidence interval version requires a choice of probability outside the confidence interval $\delta$. If the precise value for choosing $\delta$ is not well motivated, optimal variance-minimizing phase estimation version is more convenient to use~\cite{Babbush2018Encoding}.
\end{itemize}

\section{Case Study 1: Hamiltonian Simulation\label{sec:hamiltonian}}
Hamiltonian simulation implements a unitary $e^{-iHt}$ given access to a Hamiltonian $H$~\cite{low2019hamiltonian}.
This case study discusses two methods of simulating block-encoded hamiltonians: by constructing block-encodings of $\cos H$ and $\sin H$ using Chebyshev polynomials, and by using GQSP to directly implement approximations for $e^{-iHt}$.
Both these methods use \emph{qubitization} to convert the input block-encoding of $H$ into a special block-encoding with useful properties.
We first describe qubitization, and then proceed to show Qualtran implementations of the two hamiltonian simulation techniques stated above.

\subsection{Qubitization\label{subsec:qubitization}}
Qubitization~\cite{low2019hamiltonian} is a technique to transform an arbitrary block encoding $\cB[A]$,
to an equivalent block encoding $\cW[A]$ whose eigenvalues correspond to the eigenvalues of $A$.
This enables performing operations on the block encoding $\cW[A]$ to transform the eigenvalues of $A$.
This operator $\cW[A]$ is usually called a \emph{qubitized quantum walk operator}.
More generally, qubitization can be combined with QSP/GQSP~(\cref{sec:qsp}) to implement arbitrary polynomials of block-encoded matrices, called the \emph{Quantum Singular Value Transformation}~\cite{gilyen2019quantum}.
See Ref.~\cite{aws2023survey}, Sec. 10.4 for a detailed pedagogical exposition.

\begin{figure}[H]
 \centering
 \begin{subfigure}{.3\textwidth}
   \includegraphics[width=\textwidth]{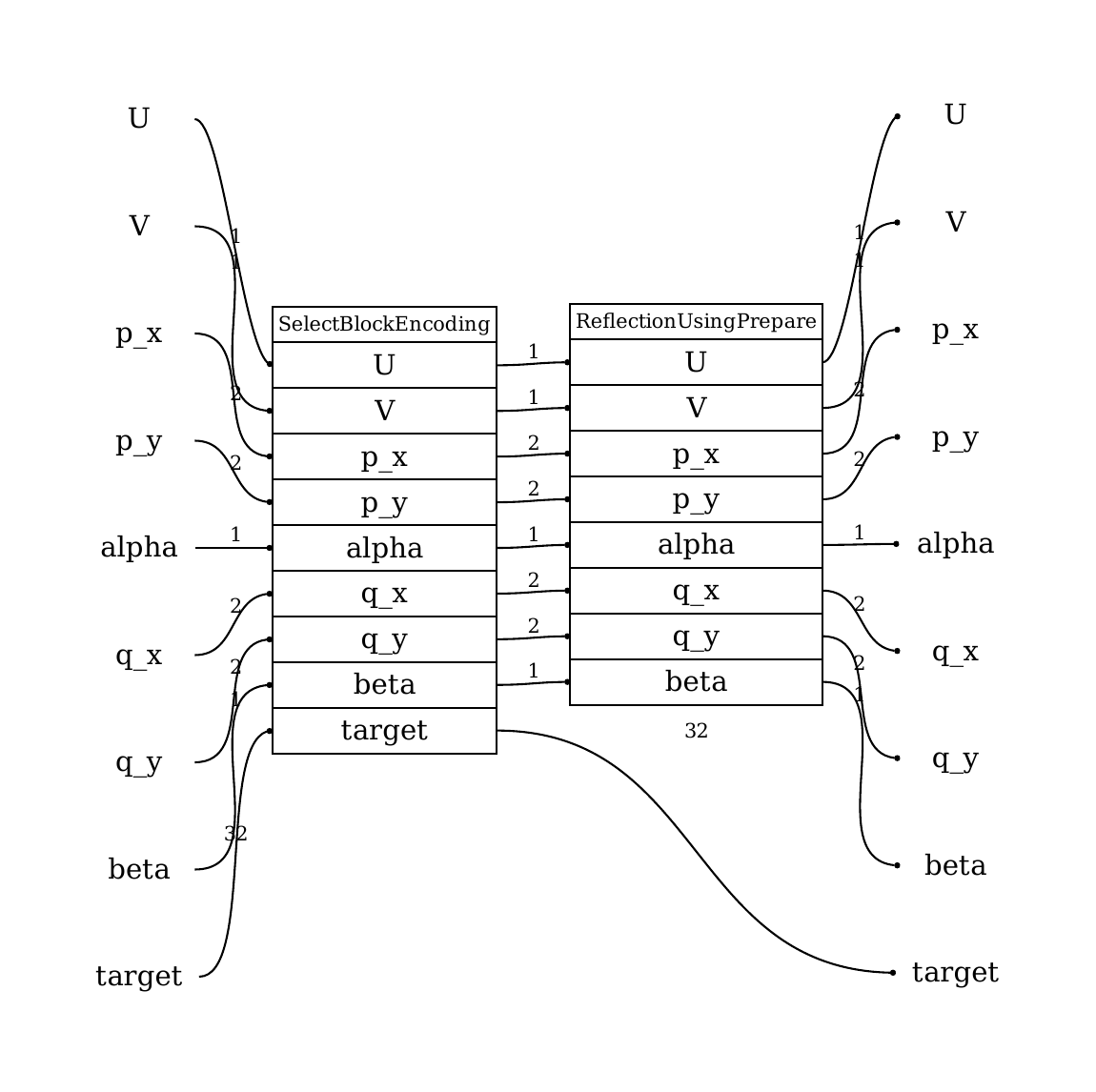}
 \end{subfigure}
 \begin{subfigure}{.69\textwidth}
   \includegraphics[width=\textwidth]{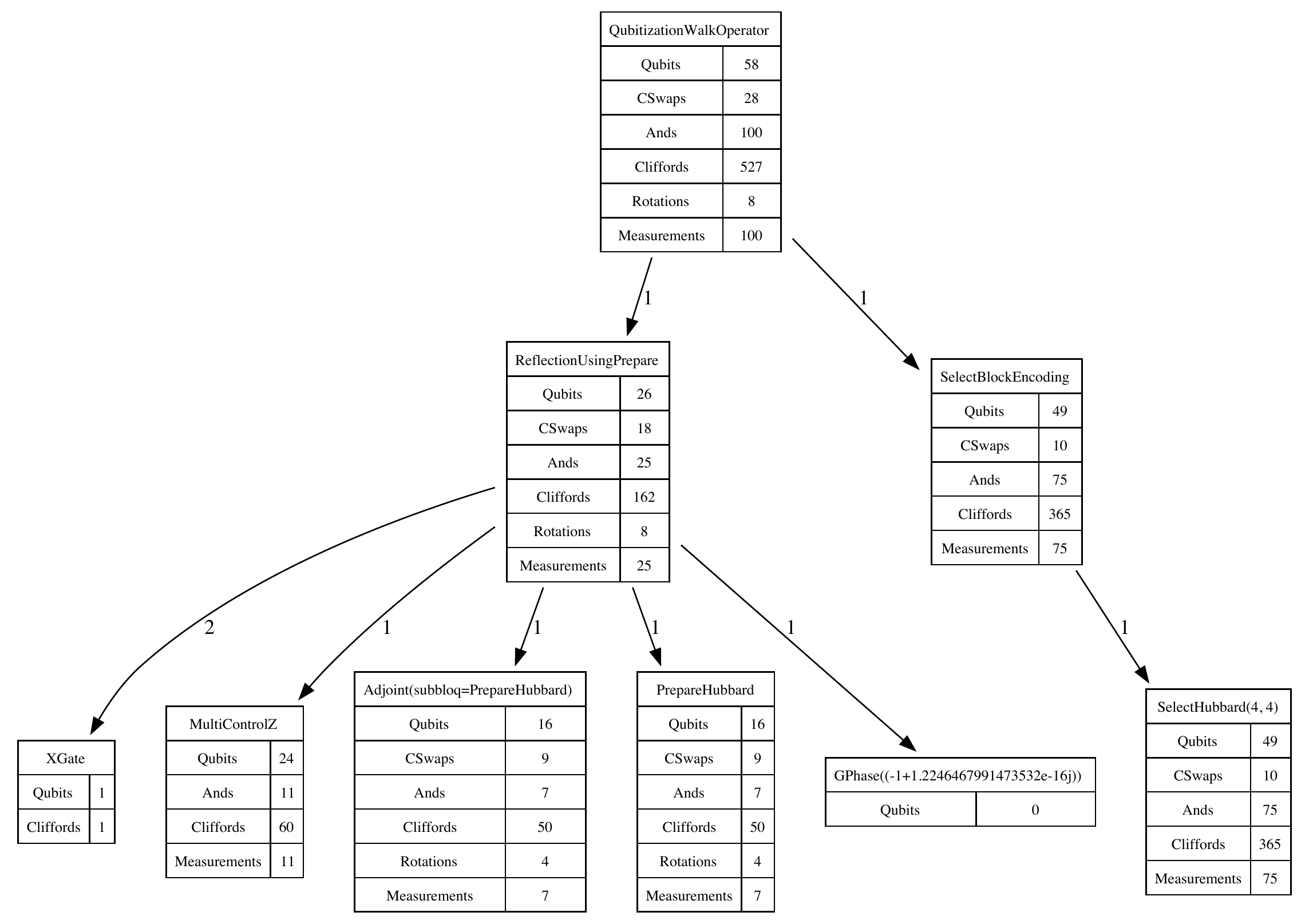}
 \end{subfigure}
   \caption{Qubitized walk operator for 2D Hubbard model following construction of \citet{Babbush2018Encoding}.}
   \label{fig:qubitization}
\end{figure}

Qualtran provides \lstinline|QubitizationWalkOperator| that implements $\cW[A]$ given a \lstinline|BlockEncoding| $\cB[A]$.
It assumes that the input block-encoding $\cB[A]$ is a reflection (i.e. $\cB[A]^2 = I$),
and implements $\cW[A]$ using one call to $\cB[A]$~\cite[Corollary 9]{low2019hamiltonian}.
\Cref{fig:qubitization} shows an example qubitized walk operator for the Hubbard Hamiltonian~\cite[Sec. 5]{Babbush2018Encoding},
using the block-encoding of the Hubbard model.

\subsection{Hamiltonian simulation using Chebyshev Polynomials}

The \lstinline`ChebyshevPolynomial` bloq uses qubitization to efficiently compute $\mathcal{B}[T_j(A/\alpha)]$, where $T_j(x)$ is the $j$-th Chebyshev polynomial of the first kind.
This uses $j$ calls to the corresponding walk operator $\cW[A]$.

The \lstinline`ScaledChebyshevPolynomial` bloq implements an appropriately scaled $\cB[T_j(A)]$,
by combining the above \lstinline`ChebyshevPolynomial` with a \lstinline`LinearCombination`.

Using the above bloqs, \cref{fig:qubitization-hamsim-chebyshev} shows an implementation of Hamiltonian simulation of a block-encoded $H$
by computing the block-encoding of $\cos(Ht)$ and $\sin(Ht)$ using Chebyshev polynomials.

\begin{figure}[H]
 \begin{subfigure}{\textwidth}
\begin{lstinputlisting}[language=Python, basicstyle=\tiny\tt]{hamsim_chebyshev.py}\end{lstinputlisting}
 \end{subfigure}
   \caption{Hamiltonian simulation using Chebyshev Polynomials}
   \label{fig:qubitization-hamsim-chebyshev}
\end{figure}

\subsection{Hamiltonian Simulation using GQSP}

\begin{figure}[H]
 \begin{subfigure}{.46\textwidth}
\begin{lstinputlisting}[language=Python, basicstyle=\tiny\tt]{snippet_hamsim_gqsp.py}\end{lstinputlisting}
 \end{subfigure}
 \begin{subfigure}{.45\textwidth}
   \includegraphics[width=\textwidth]{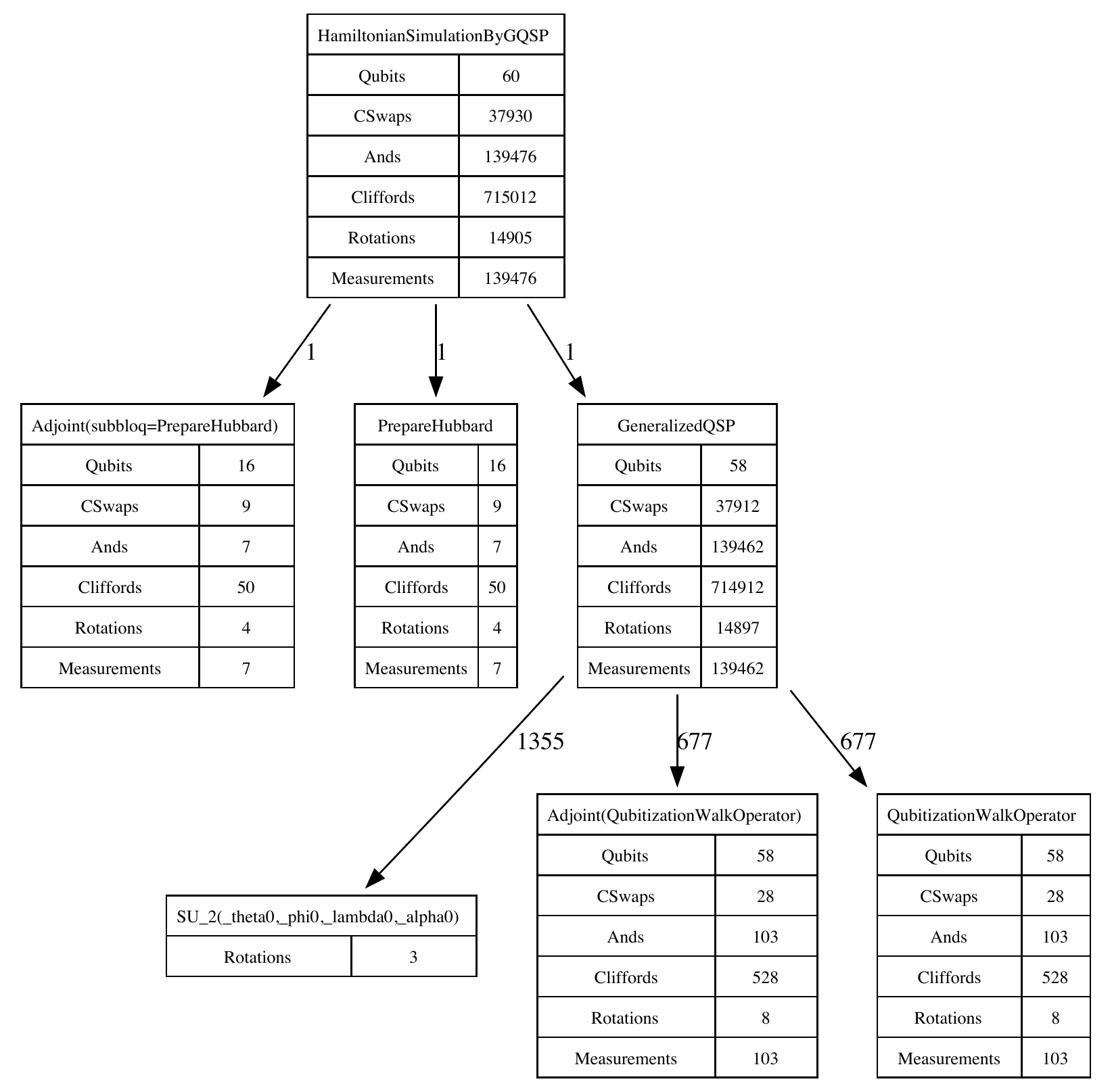}
 \end{subfigure}
   \caption{Hamiltonian simulation of 2D Hubbard using GQSP for time $t = 5$ and precision $1.0 \times 10^{-5}$, and its corresponding gate costs.}
   \label{fig:qubitization-hamsim}
 \end{figure}

\Citet{low2019hamiltonian} describe an efficient Hamiltionian simulation by combining quantum signal processing (QSP) with qubitization.
Qualtran provides the \lstinline|HamiltonianSimulationByGQSP| bloq,
which takes in a \lstinline|BlockEncoding| $\cB[H]$ which is an $(\alpha, a, 0)$-block-encoding of $H$,
a time parameter $t$,
and a precision $\epsilon$ and implements a $(1, a + 1, \epsilon)$-block-encoding of $e^{-iHt}$.
It does this by computing a GQSP polynomial $P$ that is $\epsilon$-close to $f(e^{i\theta}) = e^{-i \alpha t \cos(\theta)}$,
using the Jacobi-Anger expansion.
This GQSP bloq is applied to the qubitized walk operator of the input block-encoding.
The degree of this polynomial approximation scales as
\begin{equation*}
  \order{\alpha t + \frac{\log(1/\epsilon)}{\log\log(1/\epsilon)}}
\end{equation*}
The polynomial approximations can be found in \lstinline{qualtran.linalg.polynomial}.
\Cref{fig:qubitization-hamsim} shows an example snippet for simulating the 2D Hubbard Hamiltonian~\cite[Sec. 5]{Babbush2018Encoding} using \lstinline|HamiltonianSimulationByGQSP|.

\section{Case Study 2: Computing Ground States in Quantum Chemistry\label{sec:ground-states}}
Simulating chemical systems is one of the most cited use cases of fault tolerant quantum computers.
In particular, computing the ground state or the low lying excited states of
challenging molecules is seen as a good task for quantum phase estimation.
Over the years, the cost of performing phase estimation has reduced by 6 orders
of magnitude, from $10^{16}$ to $10^{10}$ Toffoli gates~\cite{Lee2021Even}.
Most of these improvements have been driven by the adoption of modern
factorizations~\cite{berry2019qubitization,vonburg2021catalysis,Lee2021Even} of the two electron integrals necessary to specify the Hamiltonian, 
and also through the use of different representations (i.e. first quantization~\cite{Su2021FirstQuant}.)

These algorithms can be built up from the extensive library of primitives that qualtran provides (see \cref{sec:primitives}), thus avoiding the toil of parsing 100 page papers. 
Qualtran provides block encodings for all four variants of the second quantized Hamiltonian (sparse, single-factorized, double-factorized, and tensor hypercontraction (THC)), as well as for the first quantized algorithm with and without a quantum projectile~\cite{Su2021FirstQuant, rubin2024quantum}.
Many of these algorithms are implemented in a top down fashion, without being compiled all the way down to primitive gates, while some (THC, and the sparse representation) have more detailed implementations.

\subsection{Resource Estimation and Algorithm Analysis}
\begin{figure}[h!]
\begin{subfigure}{0.6\textwidth}
\begin{lstinputlisting}[language=Python, basicstyle=\tiny\tt]{chem_sparse_be_cost.py}
\caption{Code for computing and classifying T counts for the block encoding of the sparse chemistry Hamiltonian.}
\label{fig:chem_sparse_listing_a}
\end{lstinputlisting}
\end{subfigure}
\hfill
\begin{subfigure}{0.35\textwidth}
\includegraphics[width=\textwidth]{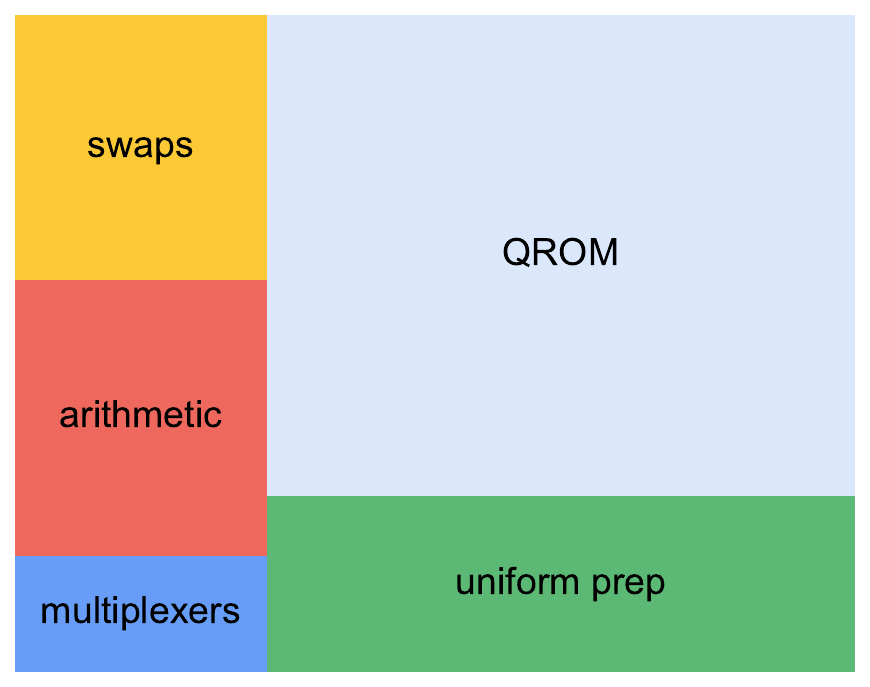}
\caption{Cost breakdown of the sparse block encoding (from \cref{fig:chem_sparse_listing_a}) in terms of different classes of primitives. Here data loading implies QROM while state preparation is a uniform state preparation.}
\label{fig:chem_sparse_listing_b}
\end{subfigure}
\caption{Example of how to generate and analyze the costs of block encodings in qualtran.}
\label{fig:chem_sparse_listing}
\end{figure}

A common task for researchers is computing the costs for performing phase estimation for a specific 
molecule of interest. 
Traditionally this has been done by  manually enumerating costs in standalone notebooks which may or may not be published, or using open source tools like OpenFermion.
While OpenFermion does provide tools for resource estimation of second quantized hamiltonians, the implemention is simply a hardcoded list of formulae with no real way to modify the values or attribute individual costs do different quantum subroutines or primitives.
Qualtran, in contrast, builds up the algorithms in a structured way, allowing for cost analysis along the call graph of the algorithm as well as enabling algorithm modification and the reuse of subcomponents.
\cref{fig:chem_sparse_listing_a} demonstrates how Qualtran provides an intuitive way to compute resource estimates and categorize the costs for a particular form of chemical Hamiltonian. 

\cref{fig:chem_sparse_listing_b} shows the utility of writing algorithms in a structured way, as qualtran can easily classify costs by the type of primitive.
Finally, \cref{tab:block_encoding_cost} shows that the block encoding costs are consistent with reported literature values (recomputed with OpenFermion for comparison) for the FeMoco model system.

\begin{figure}[h!]
    \centering
    \includegraphics[width=0.8\textwidth]{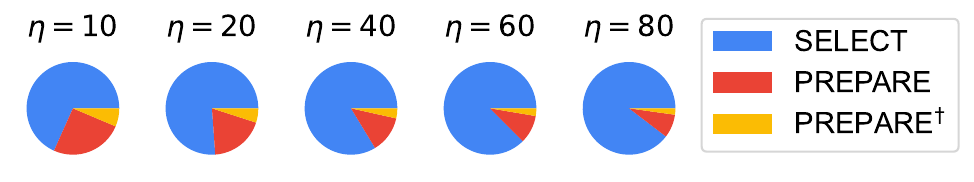}
    \caption{Breakdown in cost for the block encoding of the chemistry Hamiltonian in the first quantized representation as a function of the number of electrons $\eta$. We see that SELECT dominates which in turn is dominated by controlled swaps. PREPARE$^\dagger$ benefits from measurement based uncomputation \cite{Gidney_2018} to reduce the dominant arithmetic cost necessary for preparing the Coulomb potential $\mathrm{PREP}_\nu \propto \frac{1}{\lVert \nu \rVert^2}$. \cite{Su2021FirstQuant}   \label{fig:first_quant_sel_prep}}
\end{figure}

Qualtran also provides block encodings for the first quantized representation~\cite{Su2021FirstQuant}, which are not available in OpenFermion. 
Again, the structured way of writing algorithms in qualtran allows for additional insight into the algorithms as shown in \cref{fig:first_quant_sel_prep}, which demonstrates that in this algorithm SELECT (dominated by controlled swaps) is the dominant cost as the number of electrons ($\eta$) grows.
\begin{table}[h!]
\begin{ruledtabular}
\begin{tabular}{lccll}
& \multicolumn{2}{c}{Block Encoding} & \multicolumn{2}{c}{QPE Cost} \\
\cline{2-3}
\cline{4-5}
Method & Qualtran & OpenFermion & Qualtran & OpenFermion\\ 
\hline
Sparse & 18175 & 18011 & $4.45 \times 10^{10}$ & $4.41 \times 10^{10}$ \\
SF & 25679 & 24349 & $1.24 \times 10^{11}$ & $1.18 \times 10^{11}$ \\
DF & 34982 & 34979 & $6.44 \times 10^{10}$ & $6.44 \times 10^{10}$ \\
THC & 17150 & 16891 & $3.24 \times 10^{10}$ & $3.19 \times 10^{10}$ \\
\end{tabular}
\end{ruledtabular}
\caption{Block encoding and QPE Toffoli costs for the FeMoco model~\cite{reiher2017elucidating} computed with Qualtran compared to the reference values from~\cite{Lee2021Even} recomputed with OpenFermion~\cite{mcclean2020openfermion} \label{tab:block_encoding_cost}. The slight discrepancies arise due to some small implementation differences for some basic primitives. For example, comparators and swaps currently do not exploit measurement based uncomputation in qualtran while this optimization was used in the reference.} 
\end{table}

\subsection{Symbolic Analysis}
Quantum algorithms typically only allow us to estimate properties to within some additive error $\epsilon$, which often is a combination of different sources of errors arising from different parts of the algorithm.
An example of this is Trotterized phase estimation which introduces three sources of controllable error: 1) the Suzuki-Trotter error $\Delta_{TS}$, 2) the phase estimation error $\Delta_{PE}$, and, 3) the rotation synthesis error $\Delta_{HT}$ \cite{kivlichan2020improved,campbell2021early}.
A common task is to optimize the costs such that $\epsilon \le \Delta_{TS} + \Delta_{HT} + \Delta_{PE}$.
The rotation synthesis cost is a parameter of all rotation bloqs in qualtran, while the trotter error and phase estimation error are incorporated in higher level TrotterizedUnitary, and PhaseEstimation bloqs.
By propagating these symbolic costs, Qualtran can automatically produce mathematical expressions for the algorithm's cost. \cref{fig:sympy_trotter} provides a code-snippet for generating the Trotterized phase estimation costs for the 2D Hubbard model \cite{kivlichan2020improved,campbell2021early} while \cref{eq:trotter_costs} is the resulting output.

\begin{figure}[h!]
\begin{lstinputlisting}[language=Python, basicstyle=\footnotesize\tt]{gen_cost_expr_no_imp.py}
\end{lstinputlisting}
\caption{Code snippet for generating symbolic Trotter costs.
\label{fig:sympy_trotter}
}
\end{figure}
\begin{figure}[h!]
  \includegraphics[width=0.45\textwidth]{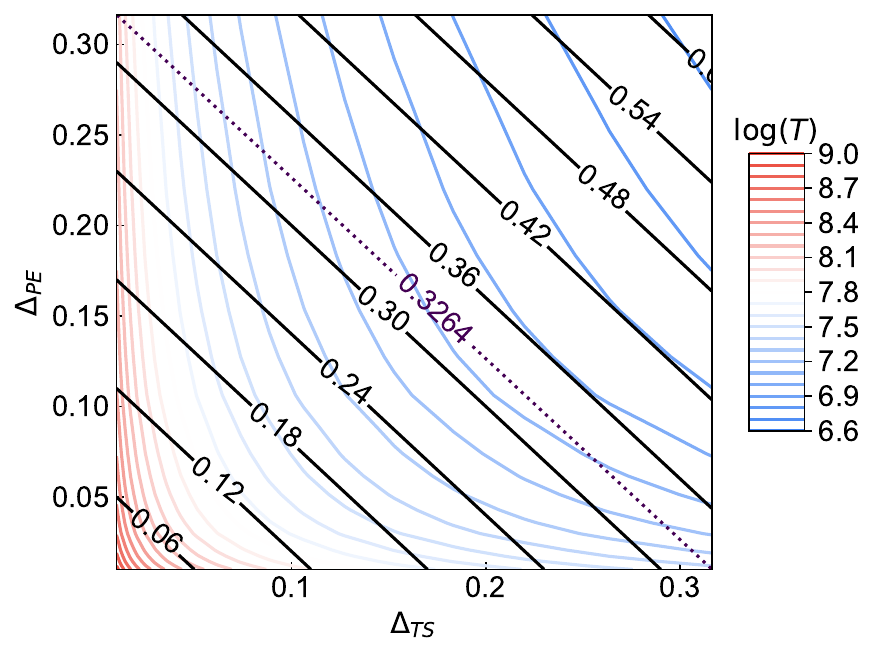}
  \caption{Contours of constant cost (\cref{eq:trotter_costs})
  as a function of $\Delta_{TS}$ and $\Delta_{PE}$, solid black lines are values
  of $\epsilon$ with the target error budget given by $\epsilon = 0.0051 L^2
  \approx 0.3264$ for $L=8$ \cite{campbell2021early}. \label{fig:trotter_costs}}
\end{figure}

\begin{equation}
T_{\mathrm{cost}} = \frac{0.76 \pi \left(\frac{\Delta_{TS}}{\xi}\right)^{- \frac{1}{p}} \left(\left(2 L^{2} + 6 \left\lfloor{\frac{L^{2}}{2}}\right\rfloor\right) \left\lceil{1.149 \operatorname{log}_{2}{\left(\frac{1.0 N_{R} \left(\frac{\Delta_{TS}}{\xi}\right)^{- \frac{1}{p}}}{\Delta_{HT}} \right)} + 9.2}\right\rceil + 24 \left\lfloor{\frac{L^{2}}{2}}\right\rfloor\right)}{\Delta_{PE}} 
\label{eq:trotter_costs}
\end{equation}
Here, $N_R$ is the number of single qubit rotations in a single Trotter step which depends on the order of the Trotter expansion $p$.
The user can numerically optimize this expression (see \cref{fig:trotter_costs}) to arrive at the lowest cost algorithm subject to the constraint on $\epsilon$. 

\section{Case Study 3: Cryptography\label{sec:cryptography}}

One of the earliest pieces of evidence for the power of quantum computation was Shor's introduction
of algorithms for factoring and the discrete logarithm~\cite{shor-1994}. With sufficiently powerful
hardware, these algorithms could be used to break commonly used cryptosystems including RSA and
elliptic curve cryptography (ECC). As such, it is particularly essential to accurately estimate the
quantum resource requirements and costs. These estimates can already impact contemporary decisions
about the urgency of the move to post-quantum cryptography e.g. for data that remains sensitive as
long as the timeframes under consideration.

Given the impact of these algorithms, a number of compilation and resource estimation studies have
been applied to optimizing the primitives used for factoring and breaking RSA and ECC
\cite{Proos2003ShorsDL,Zalka1998FastVO,Eker2017QuantumAF,Rtteler2017QuantumRE,Gouzien2023PerformanceAO,Hner2020ImprovedQC,Gidney2021howtofactorbit,litinski2023compute}.
Some studies~\cite{Rtteler2017QuantumRE, Hner2020ImprovedQC} have even published explicit code in
the Q\# programming language with complete implementations. We summarize some of the leading constructions for
RSA and ECC and show how Qualtran can be used for these algorithms. We use automated routines to highlight expensive parts of
the algorithm, show how the classical simulation protocol can aid in testing of classical reversible logic,
and automatically propagate symbolic costs for an important subroutine.

The setting for both RSA factoring and elliptic curve cryptography starts with a finite field. The
quantum algorithm uses phase estimation of the field’s group multiplication operation to extract the
hidden property of interest. The field and group operation differ between the two schemes, but the
group operation is always a classical operation that we must execute on a quantum superposition
of inputs. Optimization of these constructions, therefore, need not be performed by experts in quantum
information. The structure and organization of a library such as Qualtran can help solicit contributions
from classical computer scientists as well as traditional quantum algorithms researchers.

\begin{table}[H]
\centering
\caption{Comparison of RSA and ECC cryptographic schemes}
\def\arraystretch{1.2}
\begin{tabular}{|l|l|l|}
\hline
 & \textbf{RSA} & \textbf{ECC} \\ \hline
\textbf{Problem} & 
\begin{tabular}[c]{@{}l@{}}
$N = pq$ \\ 
private primes $p$, $q$ \\
public key $N$.
\end{tabular} & 
\begin{tabular}[c]{@{}l@{}}
$Q = [k] P \mod p$ \\ 
public key $Q$, private key $k$,\\ 
specified base point $P$ and \\
modulus $p$
\end{tabular} \\ \hline
\textbf{Discrete log} & 
\begin{tabular}[c]{@{}l@{}}
$[r] g = 1 \mod N$ \\ 
Randomly selected $g$, find \\
order $r$
\end{tabular} & 
\begin{tabular}[c]{@{}l@{}}
$Q = [k] P \mod p$ \\ 
Find private key $k$
\end{tabular} \\ \hline
\textbf{Group operation} & 
    Modular multiplication &  Elliptic curve point addition \\  \hline
\textbf{Group operation bloq} &
  \lstinline|ModMulK| &  \lstinline|ECAddR| \\ \hline 
\textbf{Typical $n = \ceil{\log_2 N}$} & 
    $2048$ & $256$ \\ \hline
\textbf{Qubit cost, approx.} & 
    $3n$ & $9n$ \\ \hline
\textbf{Toffoli cost, approx.} & 
    $0.4n^3$ \cite{Gidney2021howtofactorbit} & $43n^3$ \cite{Gouzien2023PerformanceAO} \\ \hline
\end{tabular}
\label{table:rsa_ecc}
\end{table}

For factoring large primes to break RSA encryption, the public key $N$ is the product of two hidden,
large prime numbers, $N = pq$. Using standard number theory, it suffices to find the order $r$ of
the group operation on a randomly selected element $g \in \mathbb{Z}_N$. Specifically, we seek the
order $r$ such that $[r] g = 1 \mod N$ where the notation $[r] g$ indicates repeating the group
operation $r$ times. The operation in question, and the operation on which we perform phase
estimation, is modular multiplication by the constant $g$, or, in Qualtran: \lstinline{ModMulK(k=g, mod=N)}.
In traditional mathematical notation, we would seek to find the order $g^r = 1 \mod N$.
Phase estimation takes powers of the operator, which can be classically-fast forwarded. In fact, the
phase estimation circuit is equivalent to performing the modular exponentiation operation.

In elliptic curve cryptography, the public key $Q$ is an unknown multiple $k$ of a shared base point
$P$ within a finite field of elliptic curve points with modulus $p$ such that $Q = [k] P$. 
The group operation is elliptic curve point addition and the notation $[k] P$ once again indicates
repeating the group operation $k$ times. The operation on which we perform phase estimation is
addition of a constant elliptic curve point, or, in Qualtran: \lstinline{ECAddR(R=..., mod=p)}. 
In the first phase estimation,  we set the constant to the base point $P$ to project into an eigenbasis
of the addition operation; then we phase estimate with the constant set to the public key $Q$ to extract
the hidden key. We highlight the key differences between the two encryption schemes in \cref{table:rsa_ecc},
but stress that compilation of both algorithms hinges on efficient construction of classical reversible gates.

\begin{figure}[H]
  \centering
  \begin{subfigure}{0.40\textwidth}
    \centering
    \includegraphics[scale=0.7]{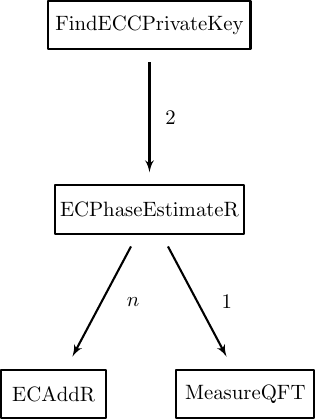}
    \caption{Naively, the algorithm consists of two phase estimations of elliptic curve point
  addition \lstinline{ECAddR} for two different, classical elliptic curve points $R$.}
    \label{fig:ecc_simple}
  \end{subfigure}
  \hspace{2em}
  \begin{subfigure}{0.40\textwidth}
    \includegraphics[scale=0.7]{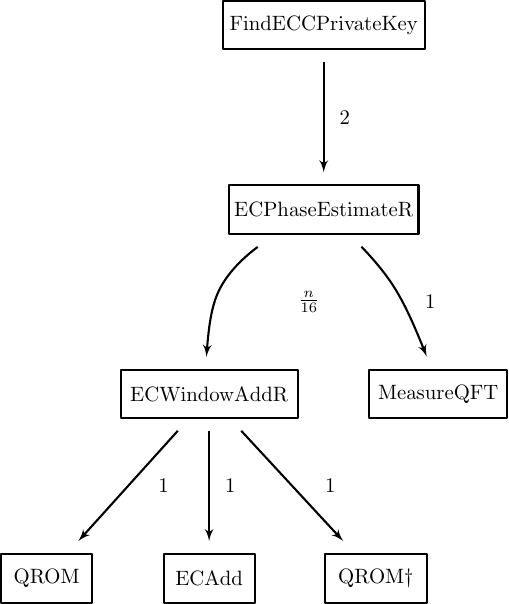}
    \caption{In practice, we use a windowing technique to reduce the number of elliptic curve point
    additions at the cost of additional classical data lookup \lstinline{QROM} operations.}
    \label{fig:ecc_window}
  \end{subfigure}
  \caption{The high-level call graph of the algorithm for finding elliptic curve cryptography private keys.}
  \label{fig:ecc_call_graphs}
\end{figure}

\subsection{Implementation in Qualtran}

For the remainder of this section, we will focus on using Qualtran to sketch out a high-level construction
of an algorithm for breaking elliptic curve cryptography following the derivation in Litinski~(2023)~\cite{litinski2023compute}. 
We find the most expensive subroutines, use the classical simulation protocol, and automatically tabulate
contributions to the cost down to basic arithmetic operations. 

The high-level call graph for the algorithm is shown in \cref{fig:ecc_call_graphs}. The algorithm consists of two
phase estimation routines. Each phase estimation ends by fourier transforming and measuring the control register.
The bulk of the computation is spent performing the $n$ applications of the group operation. Naively, we would call
the \lstinline{ECAddR} operation $n$ times for with addition constant $R = [2^j] P$ for $j = 0 \dots n-1$.
This call graph is diagrammed in \cref{fig:ecc_simple}, and the salient code for composing the point addition
operation in this way is shown in \cref{listing:ecc_simple_code}.
To reduce the cost, we can use the technique of windowed arithmetic.
In this case, we choose a window size (for example: 16 bits) and pre-compute $2^{16}$ elliptic curve points.
Each group of 16 \lstinline{ECAddR} can be replaced by a lookup operation to retrieve the pre-computed point,
a quantum-quantum \lstinline{ECAdd} operation, and an un-lookup. The call graph for this optimized scheme is shown in
\cref{fig:ecc_window}.

\begin{figure}[H]
    \begin{subfigure}{0.45\textwidth}
        \begin{lstlisting}[language=Python,basicstyle=\footnotesize\tt]
ctrl = [bb.add(PlusState()) 
        for _ in range(self.n)]
        
for i in range(self.n):
  ctrl[i], x, y = bb.add(
    ECAddR(n=self.n, 
           R=2**i * self.point), 
    ctrl=ctrl[i], x=x, y=y
  )
        
bb.add(MeasureQFT(n=self.n), x=ctrl)
return {'x': x, 'y': y}
        \end{lstlisting}
        \caption{Qualtran code for defining \lstinline{ECPhaseEstimateR} as $n$ applications
                 of \lstinline{ECAddR} with the appropriate $R$.}
        \label{listing:ecc_simple_code}
    \end{subfigure}
    \hspace{1em}
    \begin{subfigure}{0.45\textwidth}
        \begin{lstlisting}[language=Python,basicstyle=\footnotesize\tt]
P = ECPoint(15, 13, mod=17, curve_a=0)
for j in range(1, 4 + 1):
  bloq = ECAddR(n=5, R=j * P)
  ctrl, x, y = bloq.call_classically(
    ctrl=1, x=P.x, y=P.y
  )
  print(f'+[{j}] P -> ({x}, {y})')
        \end{lstlisting}
        \begin{verbatim}
+[ 1] P  ->  ( 2, 10)
+[ 2] P  ->  ( 8,  3)
+[ 3] P  ->  (12,  1)
+[ 4] P  ->  ( 6,  6)
        \end{verbatim}
    \caption{With the classical simulation protocol, we can demonstrate and test
    elliptic curve point addition on classical values.}
    \label{listing:ec_add_classical}
    \end{subfigure}
    \caption{Code snippets for composing bloqs and calling them classically.}
\end{figure}

As described in \cref{sec:classical-sim}, we can annotate bloqs with their classical action. For
many arithmetic bloqs, this can be as simple as adding two integers. But, the protocols supports
arbitrary Python code. For our point addition bloq, we introduce a Python class for encoding
elliptic curve points, define point addition and multiplication, and annotate \lstinline{ECAddR}
with that classical action. In \cref{listing:ec_add_classical}, we \emph{call} the bloq with classical
values: both the bloq attribute $R$---which is always classical for this bloq---and also
the $x$ and $y$ registers that ordinarily would contain a superposition of elliptic curve point coordinates,
but here take a classical value of $P = (15, 3)$ in this example 5-bit field. You can see the first few
multiples $[j] P$ added to the base point $P$ produced from classically calling the \lstinline{ECAddR} bloq.

\subsection{Resource Analysis}

\begin{table}[H]
\centering
\def\arraystretch{1.2}
\begin{tabular}{|l l|}
\hline
\textbf{Sub-Operation} & \textbf{Toffoli Count} \\ \hline
\lstinline|ModInv| & $6.30 \times 10^6$ \\ 
\lstinline|ModMul| & $1.50 \times 10^6$ \\
\lstinline|CModSub| & $7.17 \times 10^3$ \\ 
\lstinline|MultiCToffoli| & $4.59 \times 10^3$ \\ 
\lstinline|ModSub| & $3.07 \times 10^3$ \\ 
\lstinline|ModAdd| & $3.07 \times 10^3$ \\ 
\lstinline|CModAdd| & $2.56 \times 10^3$ \\
\lstinline|ModNeg| & $1.53 \times 10^3$ \\ 
\lstinline|ModDbl| & $1.02 \times 10^3$ \\
\lstinline|CModNeg| & $1.02 \times 10^3$ \\ \hline
\end{tabular}
\caption{Toffoli Count for the constituent operations of \lstinline|ECAdd(n=256)|. This bloq, which adds two elliptic curve points
         is called repeatedly as part of the phase estimation for doing the discrete log. Modular inversion and modular multiplication
         are the most expensive subroutines.}
\label{table:ec_add_op_counts}
\end{table}

\begin{figure}[H]
    \centering
    \includegraphics[width=1.0\linewidth]{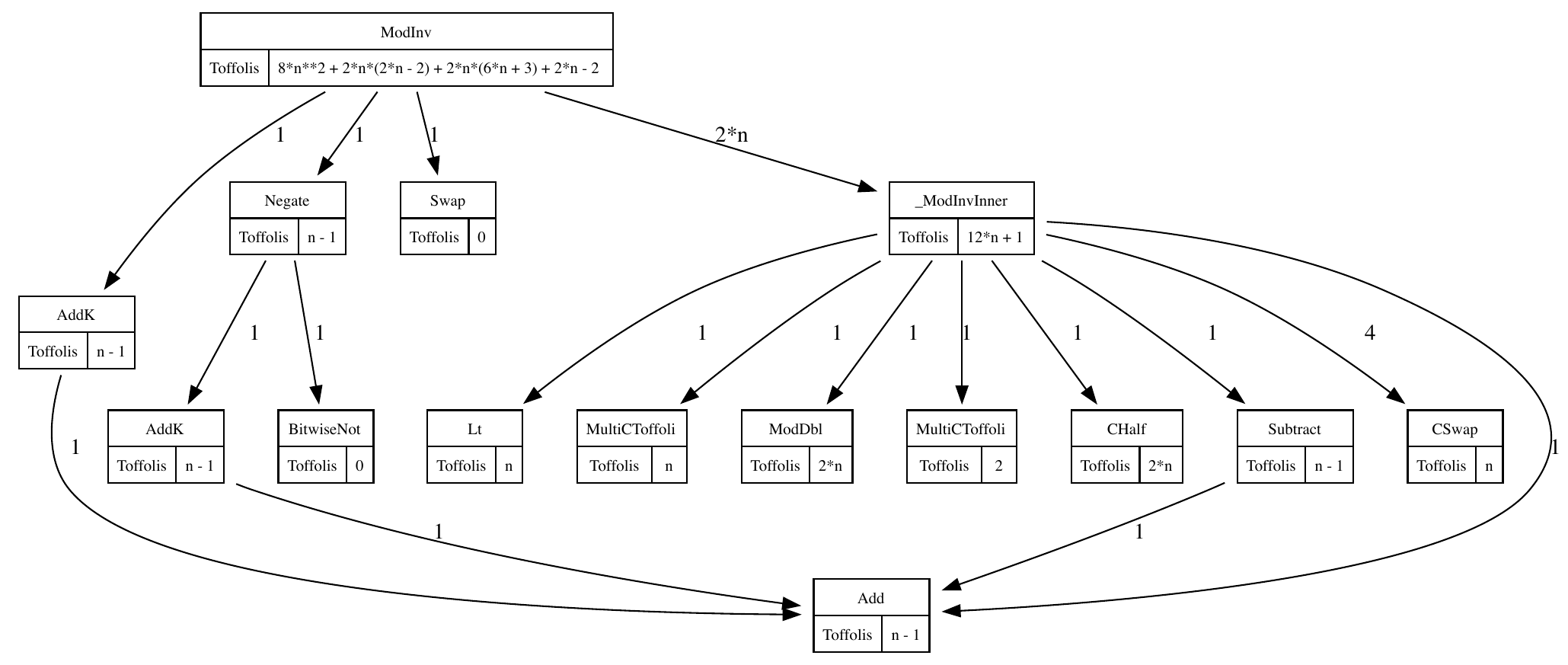}
    \caption{Symbolic gate costs for the modular inversion bloq. This call graph is based on the construction in Ref.~\cite{litinski2023compute}.}
    \label{fig:mod_inv}
\end{figure}

As we make our way down the call graph of \cref{fig:ecc_call_graphs}, we can use Qualtran
to tabulate the resource requirements for the elliptic curve point addition subroutine that
powers the algorithm. \Cref{table:ec_add_op_counts} shows the Toffoli gate count contributions from
each of the direct callees of \lstinline{ECAdd(n=256)} using the construction in Ref.~\cite{litinski2023compute}.
Modular multiplication and modular inversion are both $O(n^2)$ operations, so it is perhaps not surprising
to see that they comprise the majority of the gate cost. For the relevant choice of $n=256$, we can determine
that the \lstinline{ModInv} operation is the most costly. 

Computing a modular inverse using a reversible quantum circuit is a complicated endeavor. The circuit
implements Kaliski's algorithm~\cite{Kaliski1995TheMI}, which uses $2n$ iterations of a routine composed of comparison and arithmetic operations.
The scheme's call graph is diagrammed in \cref{fig:mod_inv}. Helpfully, Qualtran tracks the gate costs of the subroutines---here using
symbolic expressions---and propagates them through the call graph. Since the software is handling the details,
the final expression includes the linear and constant terms as well as the leading order term. We can use
explicit values for $n$ to get precise counts, or simplify the expressions to recover the leading
order modular inversion complexity of approximately $24n^2$.

\section{Analyzing Architecture-Dependent Physical Costs\label{sec:hardware-costs}}

While Qualtran provides tooling for expressing and reasoning about quantum algorithms broadly, a
particularly compelling use case for such a software tool is to provide estimates of resource
requirements for quantum algorithms that could be executed on a future quantum computer. In
particular, we’re interested in resource estimates compatible with the hardware roadmaps of quantum
hardware providers. 

A complete quantum computing system will be built with layers of
abstraction~\cite{jones2012layered}. Algorithms expressed as Qualtran bloqs are encoded at the
architecture-agnostic level \emph{application} and \emph{logical} levels of abstraction. We
anticipate that future tools or interoperability software will consume the output of a full
decomposition of a Qualtran program and explicitly compile it to a specific error-corrected
architecture or platform. In the future, Qualtran may gain support for additional, optional
annotations on bloqs to aid in architecture-specific compilation. In the meantime, we accelerate current
research practices by providing code implementations of common literature models for converting
\emph{logical costs} to \emph{physical costs}. We define these terms: logical costs are hardware
architecture-agnostic properties of an algorithm that we desire to minimize. Common logical costs
are the number of (logical) qubits or the number of (logical) gates. Physical costs are those that
matter in the physical world, such as the number of physical qubits (e.g. individual transmons or
trapped ions) and the wall-clock time required to physically execute an algorithm.

A physical cost model is encapsulated in the \lstinline{PhysicalCostModel} class and takes logical counts as inputs and outputs physical costs. The model is
parameterized by properties of the target hardware architecture
and execution protocol design choices.
The architecture properties are contained in the \lstinline{PhysicalParameters} data class.
The execution protocol's description is further factored into
the data block design for storing algorithm qubits, 
the magic state factory construction for executing gates,
and the error suppression ability of the code.
For a given model, the inputs are the number of algorithm qubits and the number of algorithm gates.
The output quantities include the wall-clock time, the number of physical qubits, and the
probability of failure due to the physical realization of the algorithm.

In Qualtran, we provide leading physical cost models for surface code architectures. The surface
code has been widely studied and is favored for only requiring 2D planar connectivity and strong
error suppression (at the expense of relatively high physical qubit costs). In particular, we
provide the \lstinline{gidney_fowler} model based on Gidney and Fowler (2019)~\cite{Gidney2019efficientmagicstate}
and related works \cite{gidney2019flexible,fowler2019low};  as well as the \lstinline{beverland} model based on Beverland~et.~al.
(2023)~\cite{beverland2022assessing}, which takes heavy inspiration from Litinksi~(2019)~\cite{Litinski2019gameofsurfacecodes}. Although we have implemented models based on the 
surface code, we welcome open-source contribution of models for alternative error correction architectures.

\subsection{Data Storage}

The physical cost model depends not only on architecture properties, but also on execution protocol
design choices. The first important design decision we consider is how to lay out the physical
qubits to support the number of logical qubits used by the
algorithm. When counting qubits, we distinguish between three distinct concepts: physical qubits,
algorithmic qubits, and tiles. 
At the highest level, we define an \emph{algorithm qubit} as a qubit
used in the routing of algorithm-relevant quantum data in a subroutine or application. 
A physical qubit is a physical system that can encode one qubit, albeit noisily.
Specifically for the surface code,
we define a \emph{tile} to be the minimal area of physical qubits necessary to encode one logical qubit
to a particular code distance $d$. A tile can store an algorithm qubit, can be used for ancillary
purposes like routing, or can be left idle. 
In the rotated surface code, it takes $2d^2$ physical qubits arranged in a grid to construct one tile.

The number of algorithm qubits is reported by Qualtran as a cost one can query on any particular
bloq. The surface code is a rate-1 code, so each bit of data needs at least one surface code tile.
Due to locality constraints imposed by the 2D surface code combined with the need to interact qubits
that are not necessarily local, additional tiles are needed to actually execute a program. This is
modeled by the \lstinline{DataBlock} abstract interface. 
Each data block is responsible for reporting the number of tiles required to store a certain number
of algorithm qubits; the expected error associated with storing a number of algorithm qubits for a
given number of cycles and a logical error model, and (optionally) the number of cycles required to
consume enough magic states to enact a given number of logical gates.

Different data block models exist in the literature. In Qualtran we provide the data blocks present
in \citet{Gidney2019efficientmagicstate} and \citet{Litinski2019gameofsurfacecodes}. The Gidney data
block model---which we term \lstinline{SimpleDataBlock}---assumes a constant 50\% overhead due to
routing and unbounded magic state consumption speed. The Litinski models provide an explicit (but
not optimal) construction for routing magic states and offers three data block designs that trade
off space (i.e.~number of tiles) and time (i.e.~number of cycles required to consume magic states).
Each data block uses $2d^2$ physical qubits per tile and has a data error that scales with the
number of tiles times the number of error correction cycles.

\subsection{Executing Gates}

The second execution protocol component that we consider is how to execute arbitrary gates
to run the desired algorithm.
The surface code can execute Clifford gates in a fault-tolerant manner.
Non-Clifford gates like the T gate, Toffoli or CCZ gate, or non-Clifford rotation gates require
more expensive gadgets to implement \cite{Horsman2011SurfaceCQ}. 
Executing a T or CCZ gate requires first using the technique of state distillation in an
area of the computation called a \emph{magic state factory} to distill a noisy T or CCZ state
into a \emph{magic state} of sufficiently low error. For example, a noisy T-state $\ket{T} = T\ket{+}$
may be created via a physical T-gate with associated physical error (e.g. $p_\mathrm{phys} = 10^{-3}$),
and then distilled down to a $\ket{T}$-state with lower error, suppressed cubically $p_\mathrm{distil} \propto p_\mathrm{phys}^3$.
Such quantum states can be used to enact the non-Clifford quantum gate through gate teleportation.
Magic state production is thought to be an important runtime and qubit-count bottleneck in 
foreseeable fault-tolerant quantum computers. With one factory per computation, the rate of magic
state production is the bottleneck. Increasing the number of magic state factories to speed
the computation trades off with the number of qubits available to form the data block, limiting the
size of problems the quantum computer can solve.

The \lstinline{MagicStateFactory} abstract interface specifies that each magic state factory must
report its required number of physical qubits, the number of error correction cycles to produce
enough magic states to enact a given number of logical gates and an error model, and the expected
error associated with generating those magic states. 

There are a handful of carefully designed magic state factories in the literature~\cite{gidney2019flexible,Gidney2019efficientmagicstate,Litinski2019magicstate}.
Magic state factories are usually defined as circuits which themselves expect to be run on a surface code. 
Note that the code distance $d$ of the tiles used in magic state factories need not be
the same as that of the data block.
A given factory construction has a prescribed physical qubit footprint and number of cycles to
produce magic states. 
The error associated with magic state production has two sources: 
data errors can occur while executing the state distillation circuit,
and distillation errors can fail to reject bad magic states from entering the computation.

\subsection{Error Correction Procedure}

Error correcting codes suppress error but do not eliminate it. The construction of the
surface code takes a parameter $d$ that scales the amount of error suppression by using more
physical qubits per tile. The exact relationship between $d$ and the resulting logical
error rate will likely be experimentally determined when sufficiently large quantum
computers are available, but in the interim we can model the relationship with an
equation of the form
\begin{equation} \label{eq:logical_error_rate}
p_l(d) = A \left(\frac{p_p}{p^*}\right)^\frac{d + 1}{2}
\end{equation}
relating the logical error rate $p_l$ to the physical error rate $p_p$
for a code distance $d$. The coefficients $A$ and $p^*$ can be fit to
numerical simulations of a surface code, but common resource estimates in the
literature use `back of the envelope' values. $p^*$ is sometimes identified with
the error threshold of the surface code. Sometimes $p^*/p_p = \Lambda$ is treated as the fit
parameter.

In Qualtran, the properties of the error correction procedure are contained within the
\lstinline{QECScheme} data class, and we provide presets for Fowler and Gidney parameters $A=0.1$, $p^*=0.01$ \cite{fowler2019low} and
Beverland parameters $A=0.03$, $p^*=0.01$ \cite{beverland2022assessing}

\subsection{Architecture Properties}

The final set of inputs for getting physical costs are the physical properties of the hardware. 
The physical error rate of unprotected (physical) operations depends on the qubit modality, 
calibration quality, and myriad other real-world considerations. We assume the physical error processes can be
sufficiently represented by an aggregate error rate $p_\mathrm{phys}$. The ratio of the physical error rate to the
$p^*$ value of the QEC scheme sets the error suppression rate and informs the choice of the code distance $d$.

The \emph{cycle time} of the hardware---i.e.~how long it takes to execute one cycle of measuring the stabilizers of the code---
sets the effective clock speed of the computation. This time parameter is the second hardware parameter. Both the physical error
rate and the cycle time are encapsulated in the \lstinline{PhysicalParameters} data class, and we provide pre-configured
choices of parameters to match values encountered in the literature \cite{Gidney2019efficientmagicstate,beverland2022assessing}
under realistic or optimistic hardware assumptions for superconducting, ion, or Majorana qubit modalities.

\subsection{Physical Cost Models}

By combining the \lstinline{DataBlock}, \lstinline{MagicStateFactory}, \lstinline{QECScheme}, and \lstinline{PhysicalParameters}
specifications, we construct a \lstinline{PhysicalCostModel} object that can produce physical qubit counts and timing information. 
In addition to pre-configured models that match complete models from the literature, we have
designed the code interfaces so each of the components can be mixed-and-matched to choose the most
relevant or compelling model, as well as highlighting the considerable variability in physical cost
estimates based on highly speculative architecture-aware assumptions.

The complete physical cost models take logical costs---the number of algorithm qubits and logical gate counts---and
return the number of cycles of computation required, the wall-clock time required to perform that many cycles, the number
of physical qubits required to store the complete computation including space for routing and magic state production, and the
final probability of an error occurring during the computation including contributions from data storage errors and unheralded
magic state distillation failures.

Often times, we would like to solve the \emph{design problem} of figuring out the QEC scheme (in
particular: the code distance $d$) or choice of factory based on a final error budget. This problem
is not invertible in general, and there are different strategies in the literature. We provide code
for two particular strategies that work with specific classes of factories and data blocks. One
approach is to perform a grid search over the data block code distance and code distances used to
implement the magic state factories to exhaustively minimize the computational volume, as in
Ref.~\cite{Gidney2019efficientmagicstate}. An alternative is to apportion the error budget in a
prescribed way: One third of the total error budget each to rotation synthesis, magic state
distillation, and algorithm qubit data storage; and solve for code distances independently for each
portion, as in Ref.~\cite{beverland2022assessing}. Users are encouraged to use
\lstinline{PhysicalCostModel} as a foundation on which other heuristics can be developed to allocate
an error budget.

\section{Closing Remarks\label{sec:conclusion}}

The pursuit of quantum computational advantage remains a driving force in quantum information
science and quantum computation. Most computational advantages in quantum computing are articulated
in terms of asymptotic costs parameterized by a problem size and error tolerance. But in order to
quantify the true magnitude of a computational advantage, overheads of quantum error correction and
constant factors associated with the algorithms themselves must be rigorously counted. Prior work
generically accounting for the overhead of modern quantum error correction determined that
super-quadratic asymptotic speedups would likely be required for a quantum runtime
advantage~\cite{Babbush_2021} further enhancing prior work that the overheads of quantum error
correction can be severe~\cite{PhysRevA.79.062314}. 

We presented Qualtran, a software framework for developing and analyzing quantum algorithms. We
demonstrated how Qualtran's features for expressing and analyzing quantum programs can be applied to
problems in Hamiltonian simulation, chemistry, and cryptography. The implementations of those
algorithms were expressed as \emph{bloqs}---our foundational data structure. The bloqs are
hierarchically composed, and the case studies were built on top of the standard library of quantum
subroutines included in the package. Features for bloq construction, handling of quantum data types,
graph traversal resource estimation routines, tensor simulation of bloqs, and physical resource
estimation can accelerate algorithm analysis.

A driving philosophy of Qualtran is that the benefits of software-based tooling can complement
and accelerate researchers' existing or forthcoming resource estimation projects.
Those benefits include making constructions explicit, shareable, reproducible, tested, and reusable.
The key idea of the framework is that quantum algorithms can be composed
in a structured compute graph where edges represent flow of quantum data. A minimal
specification of each node is required, which offers maximum flexibility when exploring
algorithmic ideas. In other words, various levels of abstraction can be used to specify an
algorithmic idea without the burden of a full compilation. This is especially useful for the type of
constant factor analysis that has recently been performed in modern quantum algorithms papers for
chemistry~\cite{vonburg2021catalysis, berry2019qubitization} where constant factor analysis and
algorithmic improvements are presented in enumerated lists of English prose.

Strong network effects are unlocked by using open, structured representations of quantum programs.
Qualtran provides a standard library of primitives necessary to construct many modern quantum
algorithms, and we aim to enable developers to quickly express new algorithms as a high-level idea with
defaults for each primitive selected appropriately and updated whenever improvements are discovered.
A particularly illustrative example is the unary iteration bloqs that can be used to construct a
wide variety of other more complicated primitives such as block encodings, QROM, QROAM,
variable-spaced QROM, and programmable gate arrays. As more bloqs are published, algorithms can use
them in their own implementation. We anticipate continued growth of the library of bloqs both hosted
in the main Qualtran repository and published by users. These libraries can grow not only in
the quantity of subroutines, but also in precision: a placeholder bloq annotated with a cost
expression or its classical-reversible action can have its full decomposition filled in by
independent contributors. 
We know that we have only scratched the surface with our
current implementations of quantum programs and subroutines, and look forward to continued
development of the library of algorithms.

While we have made judicious selections in the design of the framework, we anticipate that this is
only the beginning of the story of tooling for compiling quantum algorithms. We are clear on the
current limitations; the set of features in the present version of Qualtran is not comprehensive. We
anticipate extensions to represent classical data as part of the compute graph (e.g. as the result
of a measurement operation) and support for open systems simulation. We currently leverage SymPy for
symbolic algebra manipulations, but look to augment it with custom operations common in quantum
information. For example, the \lstinline{bitsize} operation of $\ceil{\log_2 x}$ ought to be
made particularly easy to express and manipulate symbolically. Re-write rules to manipulate and optimize
compute graphs would provide more flexibility and utility from a given library of bloqs. We have
protocols for automatically constructing the adjoint or controlled version of a bloq, but we imagine
that additional \emph{meta-bloqs} are possible. For example, a \lstinline{Conjugate} bloq that
encapsulates an $A^\dagger B A$ operation would likely be useful. More visualization tools and
interactive, graphical representations would make the tool more accessible. To further boost the
network effects inherent in quantum programming, greater interoperability and interchange with other
tools and representations is desired. The development of these features must be prioritized by the
needs of the algorithms research community. 

The proliferation of long resource estimation studies that manually (re-)compile quantum subroutines
and provide lossy references to previous manuscripts is not sustainable. For research to flourish,
open, reproducible constructions should be published. We call on the community to provide explicit
and structured representations of their subroutines alongside resource estimates; potentially via
Qualtran. Quantum algorithms research is still in its early stages: novel asymptotic speedups and
constant factor improvements remain to be discovered. By providing precise resource requirements, we can
guide hardware development roadmaps and investment decisions in the field. We hope Qualtran can play
a part in this journey and accelerate the vibrant field of finding computational advantage.

\section{Code Availability}
The Qualtran code is available at \href{https://github.com/quantumlib/Qualtran}{github.com/quantumlib/Qualtran}.

\section{Acknowledgements\label{sec:acknowledgements}}
We thank Arren Bustamante, Frankie Papa, Samuel Kushnir and 
Victor Carballo Araruna for their contributions to the Qualtran standard library. 
We thank Doug Strain and Pavol Juhas for their contributions to enhance the code quality and maintainability.  
The up-to-date list of contributors can be found on \href{https://github.com/quantumlib/Qualtran/graphs/contributors}{GitHub}.
We thank Kevin M.Obenland and the \href{https://github.com/isi-usc-edu/pyLIQTR}{pyLIQTR} team for providing valuable feedback as early adopters of the library.
We thank the Google Quantum AI team for providing an environment where this work was possible.

\bibliography{reference.bib}
\end{document}